\PassOptionsToPackage{table}{xcolor}
\documentclass[manuscript,screen,nonacm]{acmart}
\setcopyright{none}
\ifdefined\DeclareUnicodeCharacter
  \DeclareUnicodeCharacter{2192}{\ensuremath{\rightarrow}}
\fi

\usepackage{algorithm}
\usepackage{algorithmic}
\usepackage{textcomp}
\usepackage{subcaption}
\usepackage{array}
\usepackage{siunitx}
\usepackage{tikz}
\usepackage{makecell}
\usepackage{multirow}
\usepackage{multicol}
\usepackage{wrapfig}
\usepackage{needspace}
\usepackage{placeins}
\usepackage{float}
\usepackage{enumitem}
\usepackage{threeparttable}
\usepackage{xurl}
\usepackage{listings}
\usepackage[most]{tcolorbox}
\usepackage{longtable}
\usepackage{flafter}
\usepackage{graphicx}
\usepackage{tikz}
\usepackage{rotating}

\AtBeginDocument{%
}

\definecolor{myyellow}{rgb}{1,1,0.234}
\definecolor{vscodeblue}{HTML}{0000FF}
\definecolor{vscodecyan}{HTML}{267F99}
\definecolor{vscodegreen}{HTML}{008000}
\definecolor{vscodepurple}{HTML}{795E26}
\definecolor{vscodenumber}{HTML}{098658}
\definecolor{vscodegray}{HTML}{6A9955}
\definecolor{codebg}{HTML}{F8F8F8}
\definecolor{coderule}{HTML}{D6D6D6}

\lstdefinestyle{codecell}{
  language=Java,
  basicstyle=\ttfamily\scriptsize,
  keywordstyle=\bfseries\color{vscodeblue},
  commentstyle=\itshape\color{vscodegray},
  stringstyle=\color{vscodegreen},
  identifierstyle=\color{black},
  emph={body,prep},
  emphstyle=\color{vscodepurple},
  morekeywords={case,switch,break,else,if},
  backgroundcolor=\color{codebg},
  rulecolor=\color{coderule},
  frame=single,
  framerule=0.25pt,
  framesep=0pt,
  columns=fullflexible,
  keepspaces=true,
  breaklines=true,
  aboveskip=0.25em,
  belowskip=0.15em,
  xleftmargin=0pt,
  xrightmargin=0pt,
  escapeinside={(*@}{@*)},
  literate=
    {0}{{{\color{vscodenumber}0}}}{1}
    {1}{{{\color{vscodenumber}1}}}{1}
    {2}{{{\color{vscodenumber}2}}}{1}
    {3}{{{\color{vscodenumber}3}}}{1}
    {4}{{{\color{vscodenumber}4}}}{1}
    {5}{{{\color{vscodenumber}5}}}{1}
    {6}{{{\color{vscodenumber}6}}}{1}
    {7}{{{\color{vscodenumber}7}}}{1}
    {8}{{{\color{vscodenumber}8}}}{1}
    {9}{{{\color{vscodenumber}9}}}{1}
}

\newtcolorbox{resultbox}{
  colback=gray!8,
  colframe=black,
  boxrule=0.6pt,
  arc=1mm,
  left=2mm,
  right=2mm,
  top=1mm,
  bottom=1mm
}

\newcommand{\HeatRangeMin}{-0.5}
\newcommand{\HeatRangeMax}{0}

\newcommand{\HeatNum}[1]{%
  \pgfmathsetmacro{\rounded}{round(1000*(#1))/1000}%
  \makebox[4.8em][c]{%
    \ifdim \rounded pt > 0pt
      +\num[round-mode=places,round-precision=3]{#1}%
    \else
      \ifdim \rounded pt < 0pt
        \num[round-mode=places,round-precision=3]{#1}%
      \else
        =\num[round-mode=places,round-precision=3]{#1}%
      \fi
    \fi
  }%
}

\newcommand{\DisplayPrecision}{3}

\newcommand{\HeatUnitFOne}[1]{%
  \pgfmathsetmacro{\rounded}{round(10^\DisplayPrecision*(#1))/10^\DisplayPrecision}%
  \ifdim \rounded pt < 0pt
    \pgfmathsetmacro{\heatT}{max(0,min(1,(\rounded-\HeatRangeMin)/(\HeatRangeMax-\HeatRangeMin)))}%
    \pgfmathtruncatemacro{\heatLvl}{min(4,max(0,int(round(4*(1-\heatT)))))}%
    \ifcase\heatLvl
      \cellcolor{yellow!10}%
    \or
      \cellcolor{yellow!25}%
    \or
      \cellcolor{orange!35}%
    \or
      \cellcolor{red!35}%
    \or
      \cellcolor{red!60}%
    \fi
  \else
    \cellcolor{green!12}%
  \fi
  \HeatNum{#1}%
}

\begin{document}

\title{Characterizing Robustness in Semantic Code Clone Detection}

\author{Zhiwei Xu}
\makeatletter
\g@addto@macro\@authornotes{%
  \if@ACM@anonymous\else
    \begingroup
    \renewcommand\thefootnote{}%
    \let\@makefnmark\relax
    \footnotetext{Zhiwei Xu and Weixian Deng contributed equally to this work.\newline Xibin Zhao is the corresponding author.}%
    \endgroup
  \fi}
\makeatother
\orcid{0000-0003-4430-3727}
\affiliation{%
  \institution{KLISS, BNRist, School of Software, Tsinghua University}
  \city{Beijing}
  \country{China}
}
\email{xzw24@mails.tsinghua.edu.cn}

\author{Weixian Deng}
\orcid{0009-0002-9197-6457}
\affiliation{%
  \institution{KLISS, BNRist, School of Software, Tsinghua University}
  \city{Beijing}
  \country{China}
}
\email{deng-wx23@mails.tsinghua.edu.cn}

\author{Xuyang Liu}
\orcid{0009-0005-9339-9755}
\affiliation{%
  \institution{KLISS, BNRist, School of Software, Tsinghua University}
  \city{Beijing}
  \country{China}
}
\email{liuxuyan23@mails.tsinghua.edu.cn}

\author{Xiaolin Peng}
\orcid{0009-0004-6167-2318}
\affiliation{%
  \institution{KLISS, BNRist, School of Software, Tsinghua University}
  \city{Beijing}
  \country{China}
}
\email{pxl25@mails.tsinghua.edu.cn}

\author{Jiabao Gao}
\orcid{0009-0008-6544-9382}
\affiliation{%
  \institution{KLISS, BNRist, School of Software, Tsinghua University}
  \city{Beijing}
  \country{China}
}
\email{gjb25@mails.tsinghua.edu.cn}

\author{Tian Qiu}
\orcid{0000-0003-2005-1326}
\affiliation{%
  \institution{KLISS, BNRist, School of Software, Tsinghua University}
  \city{Beijing}
  \country{China}
}
\email{qt24@mails.tsinghua.edu.cn}

\author{Hai Wan}
\orcid{0000-0002-9608-5808}
\affiliation{%
  \institution{KLISS, BNRist, School of Software, Tsinghua University}
  \city{Beijing}
  \country{China}
}
\email{wanhai@tsinghua.edu.cn}

\author{Xibin Zhao}
\orcid{0000-0002-6168-7016}
\affiliation{%
  \institution{KLISS, BNRist, School of Software, Tsinghua University}
  \city{Beijing}
  \country{China}
}
\email{zxb@tsinghua.edu.cn}

\renewcommand{\shortauthors}{Xu et al.}

\begin{abstract}
Semantic code clone detection seeks to recognize shared functionality across different implementations, making robustness to behavior-preserving changes central to its effectiveness. Prior studies have exposed failures under such changes, motivating a systematic examination of how sensitivity varies with the rewrites used for evaluation. We present Systematic Code Transformation (SCT), a framework organizing 51 operators by code level, target, and edit category, and evaluate 14 token, tree, graph, and prompt-based detectors on BigCloneBench, Google Code Jam, and OJClone. By comparing transformation groups, repeated application, and alternative constructions, we connect overall performance changes with specific rewriting conditions. Of 25 token, tree, and graph detector and dataset combinations starting above 90\% F1, 21 lose at least 10 percentage points under the composition of all operators. These losses vary with transformation type and application intensity, but are not determined by edit counts or code growth alone. Even at matched locations and edit counts, alternative constructions can expose different errors despite similar aggregate losses. Further analysis distinguishes reduced score separability from changed fixed-threshold decisions with retained ranking ability, revealing different scope for threshold recovery. Overall, these findings show why benchmark effectiveness needs to extend to stable discrimination across equivalent implementations and how systematic transformation testing can assess that capability.
\end{abstract}



\maketitle

\section{Introduction}

Code clone detection identifies identical or similar code fragments and supports maintenance tasks such as locating duplicated implementations and propagating bug fixes \cite{koschke2007survey, juergens2009code, li2006cp}. In practice, developers may implement the same functionality using different expressions and control structures. These implementations form semantic clones even when their syntax and structure differ \cite{saini2018oreo, xu2024dsfm}. Recognizing this relationship requires looking beyond a program's particular implementation. When implementation details change without altering program behavior, a detector should preserve its correct judgment of the functional relationship. Behavior-preserving rewriting therefore directly tests whether detection remains effective across different implementations of the same functionality \cite{zhang2022carrot, panichella2025morph}.

Progress in code representations has made this question increasingly relevant. Learning-based detectors use lexical content, tree structures, and graph dependencies to predict clone relationships \cite{feng2024machine, xu2024dsfm, mehrotra2023rubhus}, while prompt-based methods obtain judgments through task instructions \cite{dou2023towards, zhu2025empirical}. These approaches have achieved strong benchmark results \cite{xu2024dsfm, yu2025multiple}. For example, the fine-tuned CC2Vec-3L model reports 0.96 F1 on BigCloneBench \cite{dou2024cc2vec}. These results demonstrate successful detection on the evaluated pairs. To assess how far this success extends across implementations, we examine whether the same functional relationships remain recognizable after behavior-preserving changes.

Previous research has begun to address this concern. CloneGen generates behavior-preserving variants that expose detection failures and compares the effects of transformation groups \cite{zhang2023clonegen}. ALERT studies identifier substitutions that mislead CodeBERT and GraphCodeBERT \cite{yang2022natural}, while research on previously unseen functionalities highlights broader generalization difficulties \cite{kitsios2025detecting}. These studies establish that strong original performance can coexist with sensitivity to implementation changes \cite{tian2023coda, panichella2025morph}. They also motivate a closer examination of which rewriting choices expose that sensitivity and how they shape the resulting robustness assessment. A transformation can be applied in different amounts or realized through different constructions, so the chosen rewrite may influence both the measured loss and the failures uncovered. We investigate these dimensions together across detectors and datasets, then examine prediction scores and errors to understand how the resulting losses manifest in clone discrimination.

\begin{figure}[t]
    \centering
    \includegraphics[width=\linewidth]{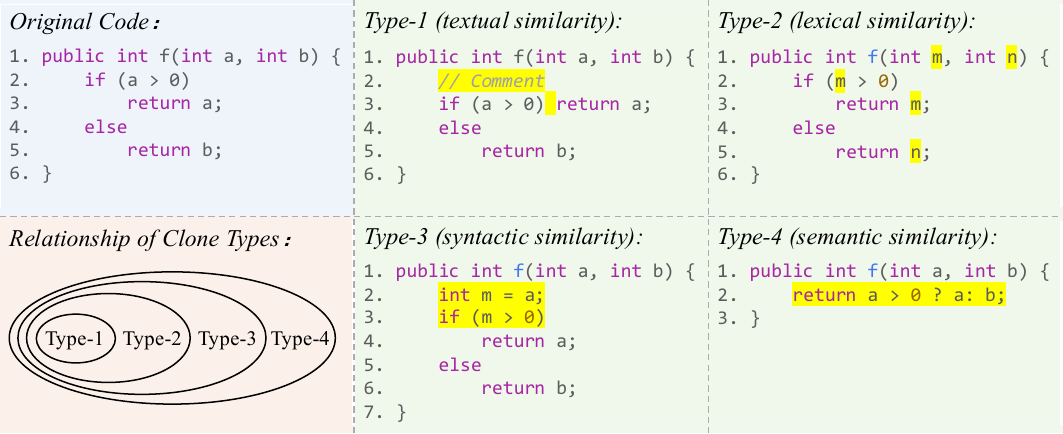}
    \caption{Examples of the four code clone types. Differences from the original code are highlighted in \colorbox{myyellow}{yellow}.}
    \label{fig:example}
\end{figure}

Making these comparisons requires varied, interpretable changes that allow detector responses to be related to concrete code modifications. The clone taxonomy in Figure~\ref{fig:example} provides a starting point by identifying lexical and structural differences between implementations \cite{saini2018oreo}. We select changes to identifiers, expressions, statements, and control structures that preserve behavior under explicit applicability conditions. By varying these identifiable elements while retaining each pair's functional relationship, we can compare detection before and after rewriting and examine whether different forms of change produce different responses. The taxonomy thus motivates the variation to investigate, while the applicability conditions govern which concrete rewrites can be used.

To organize this variation into a repeatable evaluation process, we develop \textit{Systematic Code Transformation} (SCT). SCT extends established rewriting rules with additional concrete constructions, organizing the existing rules and extensions into 51 operators classified by code level, transformation target, and edit category. Operators with the same target and category form a transformation family, and each operator specifies a concrete construction and its applicability conditions. This organization links the changes made to code with the comparisons needed for the study. Compositions examine combined changes, while repeated application reveals how responses develop as rewriting accumulates. Within a family, comparing alternative constructions at common applicable locations with matched edit counts isolates the construction choice from differences in edit location and quantity. Together, these configurations support examination of shared sensitivities and differences that a single construction could leave unobserved.

Using SCT, we evaluate 14 clone detection methods spanning token, tree, graph, and prompt-based approaches on BigCloneBench (BCB) \cite{svajlenko2014towards}, Google Code Jam (GCJ) \cite{zhao2018deepsim}, and OJClone \cite{mou2016convolutional}. This range allows us to examine whether sensitivities recur across approaches or depend on particular detectors and code changes. We begin with overall performance and prediction consistency, then examine transformation type, repeated application, and alternative constructions. Finally, error and score analyses characterize the different forms of degradation. The study addresses four research questions.

\begin{itemize}
    \item \textbf{RQ1. To what extent does strong benchmark performance translate into robustness to behavior-preserving transformations?} High original performance motivates examining whether performance advantages and correct predictions persist after rewriting.

    \item \textbf{RQ2. How does detector robustness vary with transformation type and intensity?} Different changes may expose different sensitivities. We compare transformation groups and repeated application alongside the realized rewriting amounts.

    \item \textbf{RQ3. How do alternative constructions within a transformation family affect robustness assessments?} A family can be realized through several constructions. We compare their performance losses and newly introduced errors at common locations with matched edit counts.

    \item \textbf{RQ4. How do behavior-preserving transformations affect detectors' ability to distinguish clones from non-clones?} To characterize how degradation appears in detection behavior, we examine error directions, score separability, threshold recovery, and input processing.
\end{itemize}

The results reveal a substantial gap between original effectiveness and robustness to rewriting. Of 25 token, tree, and graph detector and dataset combinations starting above 90\% F1, 21 lose at least 10 percentage points under ALL51, the composition of all operators.\footnote{We acknowledge that our prior systems, xASTNN~\cite{xu2023xastnn} and DSFM~\cite{xu2024dsfm}, also exhibit the robustness limitations examined in this study.} Detector rankings can reverse, and similar aggregate scores can conceal changes in which pairs are correctly classified. Examining the rewriting conditions shows that sensitivity varies with transformation type and repeated application, while more edits do not necessarily cause larger losses. Even after matching locations and edit counts, alternative constructions can reveal different errors, making construction choice consequential for robustness assessment. The score analyses further distinguish weakened class separation from changed fixed-threshold decisions with retained ranking ability, leaving different scope for threshold recovery. Input processing also influences how these degradations manifest. These findings connect overall robustness gaps with the rewriting choices and detection behavior that give them their particular form.

Our contributions are as follows.
\begin{itemize}
    \item \textbf{A comparative study of 14 clone detectors on three benchmarks.} We examine original and transformed performance, prediction consistency, and sensitivity to transformation type and intensity across token, tree, graph, and prompt-based approaches.

    \item \textbf{An analysis of construction dependence and discrimination degradation.} We show how alternative constructions expose different failures and characterize changes in error direction, score separability, and threshold recovery together with the influence of input processing.

    \item \textbf{A systematic framework for transformation testing.} SCT organizes 51 operators with explicit applicability conditions for composition, repeated application, and comparison of alternative constructions.

    \item \textbf{An open-source implementation of SCT.} We have released SCT at \url{https://github.com/xu-zhiwei/CloneOperator} to support reuse and further research.
\end{itemize}

\section{Background and Related Work}

\subsection{Types of Code Clones}

Code clones are commonly categorized into four types according to the differences between their implementations \cite{xu2024dsfm, saini2018oreo}.
\begin{itemize}[leftmargin=1em,labelsep=0.4em]
    \item \textbf{Type-1 (textual similarity).} Identical code fragments except for differences in comments, whitespace, and layout.

    \item \textbf{Type-2 (lexical similarity).} Code fragments with the same syntactic structure but differences in identifiers, types, or literal values, in addition to Type-1 differences.

    \item \textbf{Type-3 (syntactic similarity).} Syntactically similar code fragments with added, removed, or modified statements, in addition to Type-1 and Type-2 differences.

    \item \textbf{Type-4 (semantic similarity).} Syntactically dissimilar code fragments that implement the same functionality.
\end{itemize}

This taxonomy describes how code fragments relate in text, lexical form, structure, and functionality. Our study asks whether detectors continue to identify the original clone relationships correctly when implementation changes preserve functionality. For example, consistently renaming a local variable can change the lexical form of code without changing its functionality. If this change causes a detector to reverse a correct prediction, it reveals a lack of robustness to that implementation change. We use such behavior-preserving transformations to examine whether a detector's ability to recognize the same functionality depends on its particular expression in code.

\subsection{Code Clone Detection Approaches}

Clone detectors use different representations to capture lexical content, syntactic structure, and program dependencies. CCFinder and SourcererCC compare token representations \cite{kamiya2002ccfinder, sajnani2016sourcerercc}. DECKARD derives representations from abstract syntax trees, while CCGraph matches program dependence graphs \cite{jiang2007deckard, zou2020ccgraph}. These representations expose different properties of the code to the comparison procedure, so a rewrite can change the information used by one detector while leaving another representation unchanged.

Learning-based detectors learn representations or decision rules from code pairs \cite{li2023zc3, ding2023concord}. Toma classifies pairs using token similarity features, while CC2Vec learns representations of typed tokens through contrastive learning \cite{feng2024machine, dou2024cc2vec}. Tree models such as ASTNN and DSFM encode statement trees or interactions between subtrees \cite{zhang2019novel, xu2024dsfm}. FA-AST incorporates flow information into syntax trees for graph-based learning \cite{wang2020detecting}. Pretraining provides another way to learn code representations \cite{hou2024llmse}. CodeBERT learns from code and natural language, while GraphCodeBERT incorporates data flow \cite{feng2020codebert, guo2020graphcodebert}. Representation and learning strategy are thus complementary dimensions, with structural information available to both task-specific and pretrained models.

Prior studies also explore large language models for code clone detection by providing code pairs and task prompts to obtain clone judgments \cite{dou2023towards, zhu2025empirical}. These methods can be evaluated without updating model parameters specifically for clone detection, and existing studies examine their performance under different prompting settings. Clone detection approaches thus differ both in their code representations and in how they produce judgments, including through task-specific classifiers and task prompts.

\subsection{Evaluation of Clone Detection Effectiveness and Robustness}

Clone detection effectiveness has been evaluated from several perspectives. Wang et al. compare methods using different code representations \cite{wang2023comparison}, Kitsios et al. examine detection of clones with unseen functionalities \cite{kitsios2025detecting}, and studies of large language models assess detection under different prompts and task settings \cite{dou2023towards, zhu2025empirical}. These studies examine detection methods and their generalization, while behavior-preserving transformation tests vary implementations while retaining their original functional relationships to assess whether detection remains effective.

Code transformations have long supported detector evaluation. Roy and Cordy proposed a mutation and injection framework that generates clones from an editing taxonomy to evaluate detection tools \cite{roy2009mutation}. CloneGen combines semantics-preserving rewriting rules with search strategies to expose detection failures, compares transformation groups, and examines adversarial training \cite{zhang2023clonegen}. ALERT studies adversarial identifier substitutions that account for code naturalness, including tests of CodeBERT and GraphCodeBERT for clone detection \cite{yang2022natural}. These studies reveal sensitivity to implementation changes and show that the chosen transformations influence the failures uncovered \cite{tian2023coda, li2024aacegen}.

Building on established rewriting rules \cite{zhang2023clonegen, yang2022natural}, we extend concrete constructions and systematically organize 51 operators into groups and families by code level, transformation target, and edit category. This organization supports comparisons of transformation type, intensity, and constructions within a family, linking overall performance changes to specific rewriting conditions. We broaden the evaluation to 14 detectors, including pretrained code models and prompt-based methods, across three benchmarks to examine whether prior findings extend to newer methods and a more systematically organized operator set. Beyond this broader evaluation, we systematically characterize robustness properties that prior studies have not fully elucidated in relation to specific rewriting conditions. For construction choice, we compare alternatives within a transformation family at common applicable locations with matched edit counts, separating construction differences from differences in edit location and quantity. We compare performance losses and newly introduced error sets to determine whether alternative constructions differ in the magnitude of degradation and in the pairs they cause detectors to misclassify. For discrimination, we combine error direction, score separability, and threshold recovery to analyze how performance changes manifest in score ranking and fixed-threshold decisions, and examine the influence of input processing on evaluation results. These analyses inform transformation selection and interpretation of evaluation results.

\section{The SCT Framework}
\label{sec:operator_framework}

To compare how different types, constructions, and intensities of behavior-preserving transformations affect clone detection, we develop the Systematic Code Transformation (SCT) framework. SCT organizes 51 operators by code level, transformation target, and edit category. Operators with the same target and edit category form a transformation family. Each operator specifies a concrete way to rewrite code and the applicability conditions required to preserve program behavior. By selecting operators and setting their application intensity, SCT generates code variants under different transformation conditions for comparing detector performance.

\subsection{Organization of Code Transformations}
\label{sec:rule_organization}

\begin{table}[t]
\centering
\caption{The 51 SCT operators and schematic examples. Alternating white and gray groups mark counterparts, with paired rows for opposite directions and single rows for operators covering both. Scaling indicates continued insertion or expansion ($\checkmark$), changes limited by eligible structures ($-$), or revisiting a site ($\bigcirc$). Examples assume applicable rules and may omit auxiliary code.}
\label{tab:operators}
\begingroup
\renewcommand{\arraystretch}{1.09}
\setlength{\tabcolsep}{7pt}
\setlength{\aboverulesep}{0pt}
\setlength{\belowrulesep}{0pt}
\setlength{\cmidrulesep}{0pt}
\resizebox{\linewidth}{!}{%
\begin{tabular}{lllll@{\hspace{8pt}}cl}
\toprule
\textbf{Level} & \textbf{Target} & \textbf{Category} & \textbf{ID} & \textbf{Operator} & \textbf{Scaling} & \textbf{Example} \\
\midrule
\multirow[c]{9}{*}{Word} & \multirow[c]{1}{*}{Identifiers} & \multirow[c]{1}{*}{Replacement} & \textsf{WI1} & Identifier renaming & $\bigcirc$ & \texttt{int n=1; use(n);} $\rightarrow$ \texttt{int k=1; use(k);} \\
\cmidrule{2-7}
 & \multirow[c]{8}{*}{\makecell[cl]{Constants}} & \multirow[c]{2}{*}{Redundancy} & \cellcolor{black!12}\textsf{WC1} & \cellcolor{black!12}Neutral term insertion & \cellcolor{black!12}$\checkmark$ & \cellcolor{black!12}\texttt{5} $\rightarrow$ \texttt{(5+0)} \\
 &  &  & \cellcolor{black!12}\textsf{WC2} & \cellcolor{black!12}Neutral term removal & \cellcolor{black!12}$-$ & \cellcolor{black!12}\texttt{(5+0)} $\rightarrow$ \texttt{5} \\
\cmidrule{3-7}
 &  & \multirow[c]{6}{*}{Replacement} & \textsf{WC3} & Addition decomposition & $-$ & \texttt{5} $\rightarrow$ \texttt{(2+3)} \\
 &  &  & \textsf{WC4} & Addition composition & $-$ & \texttt{(2+3)} $\rightarrow$ \texttt{5} \\
 &  &  & \cellcolor{black!12}\textsf{WC5} & \cellcolor{black!12}Boolean negation & \cellcolor{black!12}$\checkmark$ & \cellcolor{black!12}\texttt{true} $\rightarrow$ \texttt{!false} \\
 &  &  & \cellcolor{black!12}\textsf{WC6} & \cellcolor{black!12}Negation constant reduction & \cellcolor{black!12}$-$ & \cellcolor{black!12}\texttt{!false} $\rightarrow$ \texttt{true} \\
 &  &  & \textsf{WC7} & Subtraction representation & $\checkmark$ & \texttt{5} $\rightarrow$ \texttt{(7-2)} \\
 &  &  & \textsf{WC8} & Subtraction composition & $-$ & \texttt{(7-2)} $\rightarrow$ \texttt{5} \\
\addlinespace[1pt]\midrule\addlinespace[1pt]
\multirow[c]{20}{*}{Statement} & \multirow[c]{10}{*}{\makecell[cl]{Expressions}} & \multirow[c]{1}{*}{Reordering} & \cellcolor{black!12}\textsf{SE1} & \cellcolor{black!12}Operand exchange & \cellcolor{black!12}$\bigcirc$ & \cellcolor{black!12}\texttt{a+b} $\rightarrow$ \texttt{b+a} \\
\cmidrule{3-7}
 &  & \multirow[c]{2}{*}{Redundancy} & \textsf{SE2} & Redundant wrapping & $\checkmark$ & \texttt{a} $\rightarrow$ \texttt{(a+0)} \\
 &  &  & \textsf{SE3} & Redundant unwrapping & $-$ & \texttt{(a+0)} $\rightarrow$ \texttt{a} \\
\cmidrule{3-7}
 &  & \multirow[c]{7}{*}{Replacement} & \cellcolor{black!12}\textsf{SE4} & \cellcolor{black!12}Relational rewriting & \cellcolor{black!12}$\bigcirc$ & \cellcolor{black!12}\texttt{a<b} $\rightarrow$ \texttt{b>a} \\
 &  &  & \textsf{SE5} & Increment/decrement expansion & $-$ & \texttt{a++;} $\rightarrow$ \texttt{a=a+1;} \\
 &  &  & \textsf{SE6} & Increment/decrement contraction & $-$ & \texttt{a=a+1;} $\rightarrow$ \texttt{a++;} \\
 &  &  & \cellcolor{black!12}\textsf{SE7} & \cellcolor{black!12}Compound assignment expansion & \cellcolor{black!12}$-$ & \cellcolor{black!12}\texttt{a+=b;} $\rightarrow$ \texttt{a=a+b;} \\
 &  &  & \cellcolor{black!12}\textsf{SE8} & \cellcolor{black!12}Compound assignment contraction & \cellcolor{black!12}$-$ & \cellcolor{black!12}\texttt{a=a+b;} $\rightarrow$ \texttt{a+=b;} \\
 &  &  & \textsf{SE9} & De Morgan expansion & $-$ & \texttt{!(P\&\&Q)} $\rightarrow$ \texttt{!P||!Q} \\
 &  &  & \textsf{SE10} & De Morgan contraction & $-$ & \texttt{!P||!Q} $\rightarrow$ \texttt{!(P\&\&Q)} \\
\cmidrule{2-7}
 & \multirow[c]{5}{*}{\makecell[cl]{Declarations}} & \multirow[c]{1}{*}{Reordering} & \cellcolor{black!12}\textsf{SD1} & \cellcolor{black!12}Declaration exchange & \cellcolor{black!12}$\bigcirc$ & \cellcolor{black!12}\texttt{int a=1; int b=2;} $\rightarrow$ \texttt{int b=2; int a=1;} \\
\cmidrule{3-7}
 &  & \multirow[c]{2}{*}{Redundancy} & \textsf{SD2} & Declaration insertion & $\checkmark$ & \textit{empty} $\rightarrow$ \texttt{int t=0;} \\
 &  &  & \textsf{SD3} & Declaration removal & $-$ & \texttt{int t=0;} $\rightarrow$ \textit{empty} \\
\cmidrule{3-7}
 &  & \multirow[c]{2}{*}{Replacement} & \cellcolor{black!12}\textsf{SD4} & \cellcolor{black!12}Declaration splitting & \cellcolor{black!12}$-$ & \cellcolor{black!12}\texttt{int a=1;} $\rightarrow$ \texttt{int a; a=1;} \\
 &  &  & \cellcolor{black!12}\textsf{SD5} & \cellcolor{black!12}Declaration merging & \cellcolor{black!12}$-$ & \cellcolor{black!12}\texttt{int a; a=1;} $\rightarrow$ \texttt{int a=1;} \\
\cmidrule{2-7}
 & \multirow[c]{5}{*}{\makecell[cl]{Ordinary\\statements}} & \multirow[c]{1}{*}{Reordering} & \textsf{SS1} & Computation exchange & $\bigcirc$ & \texttt{a=1; b=2;} $\rightarrow$ \texttt{b=2; a=1;} \\
\cmidrule{3-7}
 &  & \multirow[c]{2}{*}{Redundancy} & \cellcolor{black!12}\textsf{SS2} & \cellcolor{black!12}Computation insertion & \cellcolor{black!12}$\checkmark$ & \cellcolor{black!12}\textit{empty} $\rightarrow$ \texttt{int t=0; t++; t-{}-;} \\
 &  &  & \cellcolor{black!12}\textsf{SS3} & \cellcolor{black!12}Computation removal & \cellcolor{black!12}$-$ & \cellcolor{black!12}\texttt{int t=0; t++; t-{}-;} $\rightarrow$ \textit{empty} \\
\cmidrule{3-7}
 &  & \multirow[c]{2}{*}{Replacement} & \textsf{SS4} & Computation splitting & $-$ & \texttt{x=(a+1)*b;} $\rightarrow$ \texttt{int t=a+1; x=t*b;} \\
 &  &  & \textsf{SS5} & Computation merging & $-$ & \texttt{int t=a+1; x=t*b;} $\rightarrow$ \texttt{x=(a+1)*b;} \\
\addlinespace[1pt]\midrule\addlinespace[1pt]
\multirow[c]{22}{*}{Control} & \multirow[c]{14}{*}{\makecell[cl]{Conditionals}} & \multirow[c]{1}{*}{Reordering} & \cellcolor{black!12}\textsf{CC1} & \cellcolor{black!12}Case group permutation & \cellcolor{black!12}$\bigcirc$ & \cellcolor{black!12}\texttt{1:A; 2:B} $\rightarrow$ \texttt{2:B; 1:A} \\
\cmidrule{3-7}
 &  & \multirow[c]{2}{*}{Redundancy} & \textsf{CC2} & Branch insertion & $\checkmark$ & \textit{empty} $\rightarrow$ \texttt{if(false)\{int t=0;\}} \\
 &  &  & \textsf{CC3} & Branch removal & $-$ & \texttt{if(false)\{int t=0;\}} $\rightarrow$ \textit{empty} \\
\cmidrule{3-7}
 &  & \multirow[c]{11}{*}{Replacement} & \cellcolor{black!12}\textsf{CC4} & \cellcolor{black!12}if to switch (nesting) & \cellcolor{black!12}$-$ & \cellcolor{black!12}\texttt{if(x==1||x==2) B} $\rightarrow$ \texttt{switch(x)\{1:2:B\}} \\
 &  &  & \cellcolor{black!12}\textsf{CC5} & \cellcolor{black!12}switch to if (nesting) & \cellcolor{black!12}$-$ & \cellcolor{black!12}\texttt{switch(x)\{1:2:B\}} $\rightarrow$ \texttt{if(x==1||x==2) B} \\
 &  &  & \textsf{CC6} & if to switch (cloning) & $-$ & \texttt{if(x==1||x==2) B} $\rightarrow$ \texttt{switch(x)\{1:B;2:B\}} \\
 &  &  & \textsf{CC7} & switch to if (cloning) & $-$ & \texttt{switch(x)\{1:B;2:B\}} $\rightarrow$ \texttt{if(x==1) B else if(x==2) B} \\
 &  &  & \cellcolor{black!12}\textsf{CC8} & \cellcolor{black!12}Conditional nesting & \cellcolor{black!12}$-$ & \cellcolor{black!12}\texttt{if(P\&\&Q) B} $\rightarrow$ \texttt{if(P)\{if(Q) B\}} \\
 &  &  & \cellcolor{black!12}\textsf{CC9} & \cellcolor{black!12}Conditional flattening & \cellcolor{black!12}$-$ & \cellcolor{black!12}\texttt{if(P)\{if(Q) B\}} $\rightarrow$ \texttt{if(P\&\&Q) B} \\
 &  &  & \textsf{CC10} & Branch inversion & $\bigcirc$ & \texttt{if(P) A else B} $\rightarrow$ \texttt{if(!P) B else A} \\
 &  &  & \cellcolor{black!12}\textsf{CC11} & \cellcolor{black!12}Conditional expression to if & \cellcolor{black!12}$-$ & \cellcolor{black!12}\texttt{x=P?a:b;} $\rightarrow$ \texttt{if(P) x=a;} \texttt{else x=b;} \\
 &  &  & \cellcolor{black!12}\textsf{CC12} & \cellcolor{black!12}if to conditional expression & \cellcolor{black!12}$-$ & \cellcolor{black!12}\texttt{if(P) x=a;} \texttt{else x=b;} $\rightarrow$ \texttt{x=P?a:b;} \\
 &  &  & \textsf{CC13} & Disjunction branch splitting & $-$ & \texttt{if(P||Q) B} $\rightarrow$ \texttt{if(P) B} \texttt{else if(Q) B} \\
 &  &  & \textsf{CC14} & Disjunction branch merging & $-$ & \texttt{if(P) B} \texttt{else if(Q) B} $\rightarrow$ \texttt{if(P||Q) B} \\
\cmidrule{2-7}
 & \multirow[c]{8}{*}{\makecell[cl]{Loops}} & \multirow[c]{2}{*}{Redundancy} & \cellcolor{black!12}\textsf{CL1} & \cellcolor{black!12}Loop insertion & \cellcolor{black!12}$\checkmark$ & \cellcolor{black!12}\textit{empty} $\rightarrow$ \texttt{boolean z=false;} \texttt{while(z)\{\}} \\
 &  &  & \cellcolor{black!12}\textsf{CL2} & \cellcolor{black!12}Loop removal & \cellcolor{black!12}$-$ & \cellcolor{black!12}\texttt{boolean z=false;} \texttt{while(z)\{\}} $\rightarrow$ \texttt{boolean z=false;} \\
\cmidrule{3-7}
 &  & \multirow[c]{6}{*}{Replacement} & \textsf{CL3} & for to while & $-$ & \texttt{for(I;P;U) B} $\rightarrow$ \texttt{I; while(P)\{B U;\}} \\
 &  &  & \textsf{CL4} & while to for & $-$ & \texttt{while(P) B} $\rightarrow$ \texttt{for(;P;) B} \\
 &  &  & \cellcolor{black!12}\textsf{CL5} & \cellcolor{black!12}for to do-while & \cellcolor{black!12}$-$ & \cellcolor{black!12}\texttt{for(;P;) B} $\rightarrow$ \texttt{if(P) do B while(P);} \\
 &  &  & \cellcolor{black!12}\textsf{CL6} & \cellcolor{black!12}do-while to for & \cellcolor{black!12}$-$ & \cellcolor{black!12}\texttt{do B while(P);} $\rightarrow$ \texttt{for(;;)\{B} \texttt{if(!P) break;\}} \\
 &  &  & \textsf{CL7} & do-while to while & $-$ & \texttt{do B while(P);} $\rightarrow$ \texttt{while(true)\{B} \texttt{if(!P) break;\}} \\
 &  &  & \textsf{CL8} & while to do-while & $-$ & \texttt{while(P) B} $\rightarrow$ \texttt{if(P)\{do B while(P);\}} \\
\bottomrule
\end{tabular}%
}
\endgroup

\end{table}

SCT organizes local code transformations by code level, transformation target, and edit category. It distinguishes three code levels, each covering specific transformation targets. The word level covers identifiers and constants. The statement level covers expressions, declarations, and ordinary computational statements. The control level covers conditionals and loops. These levels describe the scope of a transformation rather than disjoint categories in a language grammar. Changes to method or function interfaces, interprocedural restructuring, API substitution, and alternative algorithms are outside the present scope.

Each rule is assigned to the target whose representation it directly changes. Renaming a variable inside a loop targets an identifier, while converting the loop from one form to another targets the loop. Exchanging complete independent statements targets the statement sequence, even when those statements contain control structures. Rules that change several targets through separate steps are represented as compositions of individual transformations.

For a fixed target and a specified granularity of its constituents, SCT distinguishes three mutually exclusive edit categories according to how a rule changes those constituents.

\begin{enumerate}[leftmargin=1.5em,labelsep=0.4em]
    \item \textbf{Reordering.}
    The rule changes the order of existing constituents without modifying,
    inserting, or deleting them.

    \item \textbf{Redundancy.}
    The rule inserts or deletes behaviorally redundant constituents without
    modifying or reordering the retained constituents.

    \item \textbf{Replacement.}
    The rule performs any other rewrite, including modifying constituents
    or combining changes that do not satisfy either definition above.
\end{enumerate}

For example, exchanging unchanged operands of a commutative expression is reordering, adding zero to an integer literal is redundancy, and rewriting \texttt{a < b} as \texttt{b > a} is replacement because it changes the relational operator as well as the operand order. These categories describe the form of an edit. Each rule additionally specifies the applicability conditions for preserving program behavior.

A target and an edit category define a transformation family. Its operators specify constructions that can differ in source patterns, generated structures, and applicability conditions. For example, converting \texttt{if} to \texttt{switch} can share a branch body across case labels or duplicate it for separate cases. Both constructions belong to conditional replacement but yield different code structures for detector evaluation.

Table~\ref{tab:operators} presents the 51 operators. They comprise 9 word-level, 20 statement-level, and 22 control-level operators, with 4 reordering, 12 redundancy, and 35 replacement operators. In the identifier $XYk$, $X$ denotes the code level, $Y$ the target, and $k$ the operator number within that target. Opposite transformation directions are recorded separately when they require distinct source patterns. Their counterpart relationships indicate related constructions rather than guaranteed restoration of the original source. This organization provides the transformation groups and alternative constructions used in the empirical comparisons.

\subsection{Application and Validation of Code Transformations}
\label{sec:operator_families}
\label{sec:transformation_validation}

The operator inventory describes the code changes available in SCT. Applying these changes to a particular program also requires considering the context in which they occur. Matching a source pattern identifies candidate locations, but code with the same syntactic form can differ in its bindings, types, dependencies, and control flow. Each operator therefore checks the conditions needed by its rewrite in the current source before applying the change.

Loop conversion illustrates how this context affects a seemingly straightforward rewrite. Converting a \texttt{for} loop to a \texttt{while} loop involves moving the initialization before the loop and the update to the end of its body. This preserves the ordinary iteration path, but a \texttt{continue} targeting the original loop introduces another path to consider. In the \texttt{for} loop, that path executes the update before the next condition check. Leaving the \texttt{continue} unchanged in the resulting \texttt{while} loop would skip the update. The construction therefore needs to preserve the operations along this path as well as the transfer destination, and excludes candidates whose control flow it cannot handle. The same principle applies to other operators, with checks tailored to the relationships affected by each rewrite, such as bindings for renaming and dependencies and evaluation effects for reordering.

These conditions guide where a rewrite can be applied, while checks on the generated code provide further evidence of its validity. After each rewrite, the generator reparses the edited source and rejects candidates that fail its checks. For complete programs, validation extends to compilation and differential execution. The original and transformed versions are compiled under the same configuration and run on the same inputs, comparing exit status, standard output, and written files. Inputs are retained when the original terminates normally with stable observations, and compilation failures and behavioral discrepancies are recorded separately. The available program context determines how far this validation can proceed. Fragments lacking their enclosing definitions and dependencies are checked by parsing rather than independent execution. Applicability checks and validation thus complement each other. The former check the conditions required by a rewrite, while the latter check the syntactic validity of the generated code and the behavioral agreement of executable programs on the test inputs.

\subsection{Generation of Code Variants}
\label{sec:variant_generation}

Building on these applicability conditions and validation mechanisms, SCT generates code variants for robustness evaluation by configuring operator combinations and application intensity. Generation starts from the original source and follows the selected configuration and random seed, without consulting detector predictions.
An individual operator targets a particular code pattern and is restricted by its applicability conditions, while different implementations can contain several kinds of variation together. SCT therefore composes operators on the same source fragment to examine robustness under combined changes. Composing operators within a transformation group also covers different eligible patterns in that group, reducing dependence on the applicability of any single rule.

To compose these changes, SCT applies operators in a given order to the current source. Earlier edits can alter the locations and applicability of later ones, so the generator maintains candidate locations and handles conflicts as rewriting proceeds. Algorithm~\ref{alg:rq1_composition} specifies this process. Location updates and applicability checks keep pending edits consistent with the current source. The generator uses the counterpart relationships in Table~\ref{tab:operators} to control opposing rewrites. After each accepted edit, it records the operator and its output regions. A subsequent edit is skipped if its operator is the counterpart of the recorded operator and its footprint overlaps those regions. For competing constructions at the same location, the generator assigns a construction in advance and checks region conflicts among conversions in the same exclusive group. Source history rejects edits that return the entire source to a previously accepted state. The insertion stage uses a shared snapshot of eligible locations so that newly inserted code does not recursively create insertion opportunities within that stage. This bounds the insertion process while allowing edits at the locations available when the stage begins.

\begin{algorithm}[ht]
\caption{Composition of SCT operators for one source fragment.}
\label{alg:rq1_composition}
\small
\begin{algorithmic}[1]
\REQUIRE Source $b$, source ID $u$, ordered operators $\mathcal{O}$, seed $s$
\ENSURE Final variant $v$ and accepted-edit trace $\tau$
\STATE $v\gets b$; $H\gets\{\operatorname{hash}(b)\}$; guards $G\gets\varnothing$; $\tau\gets[\,]$
\FOR{each operator $o$ in $\mathcal{O}$}
  \IF{$o$ begins the insertion phase}
    \STATE Snapshot locations for the selected insertion operators on $v$
  \ENDIF
  \IF{$o$ belongs to an uninitialized exclusive conversion group}
    \STATE Reserve its current sites among safe constructions using $(s,u)$
  \ENDIF
  \STATE $L\gets$ distinct snapshot locations of $o$ (shared snapshot for insertions)
  \STATE Order $L$ by increasing footprint size with ties seeded by $(s,u,o)$
  \WHILE{$L\ne\varnothing$}
    \STATE Remove the next remapped location $\ell$ from $L$
    \IF{$\ell$ is invalidated or reserved for another operator, or overlaps a region in $G$ recorded for $o$'s counterpart or an operator in the same exclusive group}
      \STATE Skip $\ell$ and continue
    \ENDIF
    \STATE Recheck $o$'s applicability at $\ell$ in the current source $v$ and skip if it fails
    \STATE Attempt one location-level edit, yielding candidate $v'$ and edit record $e$
    \IF{edit accepted, $v'\ne v$, and $\operatorname{hash}(v')\notin H$}
      \STATE Remap $L$, shared snapshots, reservations, and $G$ through $e$
      \STATE Add $(o,\operatorname{outputRegions}(e))$ to $G$; append $e$ to $\tau$
      \STATE $v\gets v'$; $H\gets H\cup\{\operatorname{hash}(v)\}$
    \ENDIF
  \ENDWHILE
\ENDFOR
\RETURN $(v,\tau)$
\end{algorithmic}
\end{algorithm}

Algorithm~\ref{alg:rq1_composition} takes an ordered operator set as input, making execution order part of the generation configuration. This order affects both the changes retained in a variant and the locations available to later operators. The full composition follows the seven-phase schedule in Table~\ref{tab:rq1_schedule}, placing deletion and contraction before rearrangement and structural conversion, followed by expansion, renaming, and insertion. Performing removal operations before their expanding or inserting counterparts reduces direct cancellation of newly introduced code. It can also remove opportunities for later edits. Deleting a redundant branch, for example, removes the locations within it that could otherwise be expanded or receive insertions. Conversely, structural conversion can expose new patterns to later operators. The schedule thus defines a particular sequence of rewriting opportunities, with applicability determined by the evolving source.

\begin{table}[htbp]
\centering
\caption{Execution order of the 51 SCT operators (IDs in Table~\ref{tab:operators}), each visiting a finite snapshot of eligible locations.}
\label{tab:rq1_schedule}
\small
\setlength{\tabcolsep}{9pt}
\begin{tabular}{clp{0.63\linewidth}}
\toprule
\textbf{Phase} & \textbf{Operation} & \textbf{Operator sequence} \\
\midrule
1 & Deletion & WC2, SE3, SD3, SS3, CC3, CL2 \\
2 & Contraction & WC4, WC6, WC8, SE6, SE8, SE10, SD5, SS5, CC9, CC12, CC14 \\
3 & Rearrangement & SE1, SE4, SD1, SS1, CC1, CC10 \\
4 & Structural conversion & SD4, SS4, CC4, CC5, CC6, CC7, CC8, CC11, CC13, CL3, CL4, CL5, CL6, CL7, CL8 \\
5 & Expansion & WC3, WC5, WC7, SE5, SE7, SE9 \\
6 & Renaming & WI1 \\
7 & Insertion & WC1, SE2, SD2, SS2, CC2, CL1 \\
\bottomrule
\end{tabular}
\end{table}

Robustness evaluation concerns the extent of rewriting as well as its type. To examine this aspect, SCT applies an individual operator repeatedly, producing a trajectory of variants. Each round starts from the preceding variant and, within a source-size limit, attempts one application at each tracked original location, skipping locations that no longer satisfy the applicability conditions. Newly created statements and branches do not become additional tracked locations. Some operators extend earlier changes, while others exhaust their eligible patterns or revisit an existing representation. A round therefore specifies an opportunity to apply the operator rather than a fixed amount of change. The generator records successful applications and source lengths at each round to relate detection results to the realized edits. An application can modify several source spans, so counts and textual changes describe different aspects of intensity.

\section{Study Design}
\label{sec:study_design}

We investigate clone detector robustness under behavior-preserving transformations, examining performance changes, sensitivity to rewriting choices, and failure patterns. This section presents the design of this investigation.

\subsection{Evaluated Detectors}
\label{sec:evaluated_detectors}

We select 14 clone detection methods spanning token-based, tree-based, graph-based, and prompt-based detectors. The selection covers feature-based classification, structural modeling, contrastive learning, pretrained code models, and general-purpose language models. It combines established methods with recent advances to provide a broad basis for investigating robustness across detection approaches.

\textbf{Token-based detectors.}
We include Toma~\cite{feng2024machine}, CC2Vec~\cite{dou2024cc2vec}, CodeBERT~\cite{feng2020codebert}, and Mamba. Toma combines token features with a random forest classifier. CC2Vec learns representations of typed tokens through contrastive learning. CodeBERT uses pretrained code representations adapted for clone classification. Mamba models code token sequences with a state space architecture~\cite{dao2024mamba2} and adapts pretrained models for clone classification~\cite{liu2025can}.

\textbf{Tree-based detectors.}
We include ASTNN~\cite{zhang2019novel}, xASTNN~\cite{xu2023xastnn}, DSFM~\cite{xu2024dsfm}, and MRT-OAST~\cite{yu2025multiple}. These detectors encode abstract syntax trees and differ in their tree decomposition, subtree encoding, and modeling of interactions between code fragments.

\textbf{Graph-based detectors.}
We include FA-AST~\cite{wang2020detecting} and GraphCodeBERT~\cite{guo2020graphcodebert}. FA-AST augments abstract syntax trees with control and data flow edges. GraphCodeBERT incorporates data flow into pretrained code representations.

\textbf{Prompt-based detectors.}
Prompt-based detectors extend the study to clone judgments made through task instructions without task-specific parameter updates. We include Llama-3.1-8B-Instruct~\cite{dubey2024llama3}, GPT-OSS-20B~\cite{openai2025gptoss}, DeepSeek-V4.1-Flash-1017 (552B backbone parameters)~\cite{deepseek2026v41}, and Jev~\cite{typesafe2026jev}. The first three generate textual answers, while Jev returns a structured decision score, extending the evaluation to both text generation and direct decision outputs.

\subsection{Datasets}
\label{sec:study_datasets}

We select three widely used benchmarks for code clone detection, BCB, GCJ, and OJClone, covering Java and C programs at both method and complete-program granularity. Each example pairs two code fragments with a binary clone label.

\textbf{BCB.}
BCB contains Java methods drawn from software projects~\cite{svajlenko2014towards,clonebenchrepository}. We retain the pair identifiers and clone labels supplied by the evaluated benchmark subsets. These method fragments lack the enclosing classes and dependencies needed for isolated execution.

\textbf{GCJ.}
GCJ contains Java solutions to programming contest problems, following the corpus used in DeepSim~\cite{zhao2018deepsim,deepsimartifact}. Pairs are labeled according to whether the programs solve the same problem. The complete programs also support compilation and differential execution of transformed variants.

\textbf{OJClone.}
OJClone is derived from the corpus of programming solutions introduced by Mou et al.~\cite{mou2016convolutional} and used for clone detection by Wei and Li~\cite{wei2017supervised}. Clone labels are determined by problem identity. Its complete programs extend the evaluation beyond Java, with the evaluated implementations processed through the C frontend.

\subsection{Evaluation Metrics}
\label{sec:study_metrics}

We measure detection performance using precision, recall, and F1 on labeled clone and non-clone pairs, defined as
\begin{equation}
P=\frac{TP}{TP+FP}, \qquad
R=\frac{TP}{TP+FN}, \qquad
F1=\frac{2TP}{2TP+FP+FN},
\end{equation}
where $TP$, $FP$, and $FN$ denote true positives, false positives, and false negatives, respectively. For detector $m$, transformation configuration $c$, and evaluation set $S$, we measure the performance change as
\begin{equation}
\Delta F1_{m,c}(S)=F1_m(T_c(S))-F1_m(S),
\end{equation}
where $T_c(S)$ is the transformed evaluation set with pair identities and labels preserved. F1 is computed on $[0,1]$, and its changes are reported in percentage points, with negative values indicating degradation.

\begin{figure}[htbp]
    \centering
    \includegraphics[width=\linewidth]{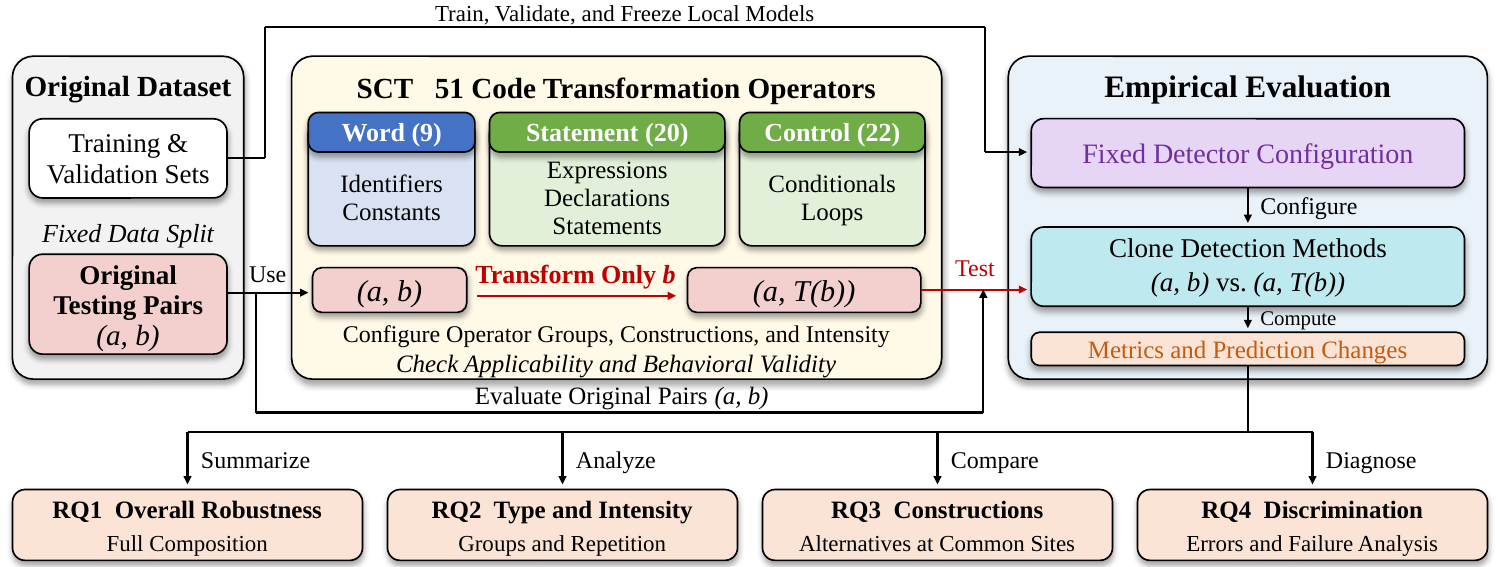}
    \caption{Evaluation workflow for comparing detector performance on original and transformed code pairs.}
    \label{fig:workflow}
\end{figure}

\subsection{Experimental Design}
\label{sec:study_protocol}
\label{sec:rq_experiments}

\subsubsection{Evaluation Workflow}

Figure~\ref{fig:workflow} presents the evaluation workflow. We first obtain predictions for the original test pairs using fixed detectors. SCT then transforms the designated second fragment of each pair, changing $(p_i,q_i)$ to $(p_i,T(q_i))$, and checks the generated variants using the validation procedure in Section~\ref{sec:transformation_validation}. We evaluate the resulting pairs with the same detectors and compare their performance and predictions. Training and validation code remain unchanged, and pairs without an accepted edit retain their original source. Full-set comparisons include these unchanged pairs, while affected-pair comparisons use the same changed pairs before and after transformation. Both comparisons retain samples successfully processed in both conditions and report their coverage.

\Needspace{250pt}
\begingroup
\setlength{\intextsep}{0pt}
\setlength{\columnsep}{0.03\linewidth}
\begin{wraptable}{r}{0.45\linewidth}
\captionsetup{font=small,skip=6pt}
\caption{Evaluation settings across datasets. O denotes the original setting, and S denotes evaluation on shared data.}
\label{tab:detector_partitions}
\small
\linespread{1}\selectfont
\setlength{\aboverulesep}{1.5pt}
\setlength{\belowrulesep}{1.5pt}
\setlength{\tabcolsep}{3pt}
\renewcommand{\arraystretch}{1.0}
\begin{tabular*}{\linewidth}{@{\hspace{\tabcolsep}\extracolsep{\fill}}llccc}
\toprule
\multirow{2}{*}[-2pt]{\textbf{Type}} & \multirow{2}{*}[-2pt]{\textbf{Method}} & \multicolumn{3}{c}{\textbf{Dataset}} \\
\cmidrule(lr){3-5}
 & & \textbf{BCB} & \textbf{GCJ} & \textbf{OJClone} \\
\midrule
\multirow{4}{*}{Token} & Toma & O & S & S \\
 & CC2Vec & S & S & S \\
 & CodeBERT & S & S & S \\
 & Mamba & S & S & S \\
\midrule
\multirow{4}{*}{Tree} & ASTNN & O & S & O \\
 & xASTNN & O & S & O \\
 & DSFM & O & O & O \\
 & MRT-OAST & O & O & O \\
\midrule
\multirow{2}{*}{Graph} & FA-AST & O & O & S \\
 & GraphCodeBERT & S & S & S \\
\midrule
\multirow{4}{*}{Prompt} & Llama-3.1-8B & S & S & S \\
 & GPT-OSS-20B & S & S & S \\
 & DeepSeek-Flash & S & S & S \\
 & Jev & S & S & S \\
\bottomrule
\end{tabular*}
\end{wraptable}

\subsubsection{Experimental Configuration}
To examine whether strong benchmark performance generalizes to behavior-preserving code transformations, we use each detector's original implementation and settings to establish its benchmark performance before applying the transformations (Table~\ref{tab:detector_partitions}). This choice is informed by our empirical observation that changes to benchmark construction may substantially reduce some detectors' performance. When a method lacks prior evaluation or a public implementation for a dataset, we evaluate it on shared data. We refer readers to the corresponding detector papers for details of dataset construction and splits in the original settings. The shared BCB and OJClone partitions follow CodeXGLUE~\cite{lu2021codexglue,microsoftcodexgluebcb} and ASTNN~\cite{zhang2019novel,astnnartifact}, respectively, while the shared GCJ partition uses the DeepSim corpus~\cite{zhao2018deepsim,deepsimartifact}.
Local models use training data for fitting and validation data for model selection. Checkpoints, preprocessing, prompts, and thresholds remain fixed across conditions. For prompt-based models, we adopt the improved simple prompt of Moumoula et al.~\cite{moumoula2025struggles}, which yielded substantial F1 gains over a basic zero-shot prompt in their evaluation. The instruction is ``Consider the overall structure and logic of the following two codes and determine if the two code snippets perform a similar task. Respond with `yes' if the two codes perform similar tasks or `no' otherwise.'' We append the code fragments under \texttt{Function A} and \texttt{Function B} and classify the pair from the final yes/no answer. Jev uses the same task instruction with a decision-score threshold of 0.5.

\par\endgroup

With detector settings fixed, we generate variants following Section~\ref{sec:variant_generation} and reuse each program's variants across evaluations under the same transformation configuration. Since individual operators apply to specific code patterns, we apply the composition of all 51 operators (ALL51) to all designated programs across the evaluated dataset partitions and validate the resulting variants. Following Section~\ref{sec:transformation_validation}, we check BCB fragments by parsing and extend validation to compilation and differential execution for complete GCJ and OJClone programs. Table~\ref{tab:validity} reports coverage and validation results. Every checked variant whose original compiles passes compilation, and differential execution detects no behavioral differences on valid test inputs.

\begin{table}[htbp]
\centering
\caption{Coverage and validation of ALL51. Compilation checks variants with compilable originals, with OJClone restricted to C programs in the shared partition. Execution uses all 557 compilable GCJ programs and 1,000 sampled OJClone programs. Six inputs are generated per program, retaining those with normal termination and stable original behavior. Compilation and execution are not applicable to BCB (shown as dashes) because its Java fragments lack the dependencies required for standalone compilation.}
\label{tab:validity}
\small
\setlength{\tabcolsep}{7pt}
\renewcommand{\arraystretch}{1.0}
\begin{tabular*}{\linewidth}{l@{\extracolsep{\fill}}ccccccccc}
\toprule
\multirow{2}{*}[-2pt]{\textbf{Dataset}} & \multicolumn{3}{c}{\textbf{Program coverage}} & \multicolumn{2}{c}{\textbf{Compilation}} & \multicolumn{4}{c}{\textbf{Differential execution}} \\
\cmidrule(lr){2-4}\cmidrule(lr){5-6}\cmidrule(lr){7-10}
 & \textbf{Changed} & \textbf{Total} & \textbf{\%} & \textbf{Passed} & \textbf{Checked} & \textbf{Selected} & \textbf{Compared} & \textbf{Valid inputs} & \textbf{Differences} \\
\midrule
BCB & 5,742 & 5,864 & 97.9 & - & - & - & - & - & - \\
GCJ & 756 & 758 & 99.7 & 557 & 557 & 557 & 383 & 1,821 & 0 \\
OJClone & 5,363 & 5,691 & 94.2 & 2,655 & 2,655 & 1,000 & 782 & 3,461 & 0 \\
\bottomrule
\end{tabular*}
\end{table}
\FloatBarrier

\subsubsection{Research Approach}

We address the four research questions through comparisons of operator compositions, transformation groups and intensity, alternative constructions, and detection failures.

\textbf{Overall robustness (RQ1).}
To test whether benchmark performance persists after rewriting, we compare detection results on original code and variants generated by LIGHT11 and ALL51. LIGHT11 contains 11 representative operators spanning all code levels, transformation targets, and edit categories (WI1, WC1, WC3, SE4, SE7, SD1, SD2, SD4, SS4, CC10, and CL4). ALL51 contains all 51 operators. We report results on both the full test set and affected pairs to distinguish overall changes from losses among pairs whose code changes. Evaluations with multiple transformation generation seeds assess the stability of these results.

\textbf{Transformation type and intensity (RQ2).}
To identify sensitivity to particular changes, we compare 13 operator groups defined by three code levels, seven transformation targets, and three edit categories. Coverage, edit counts, and code length ratios characterize the changes realized by each group. We then apply each of the eight Growth operators over multiple rounds to examine performance as rewriting accumulates. Each round makes one safe edit at each original site that remains applicable, and inserted code creates no new edit sites.

\textbf{Alternative constructions (RQ3).}
To distinguish the effect of rewritten content from edit location and quantity, we apply alternative constructions at common original sites with matched realized edit counts. The comparisons cover five pairs of alternative constructions for conditional and loop conversions. Each variant is generated independently from the original source. We compare the sets of newly misclassified pairs to determine whether similar performance losses arise from failures on the same pairs.

\textbf{Discrimination and failure analysis (RQ4).}
To distinguish changes in discrimination from threshold and input processing effects, we analyze missed clones, false positives, and prediction scores. For detectors that provide continuous scores, ROC AUC and the maximum F1 over thresholds indicate whether discrimination deteriorates and whether threshold adjustment can mitigate the losses. The maximum F1 uses test labels and serves as a diagnostic upper bound. Controlled comparisons then examine whether input truncation and batch context introduce additional detection errors.

\section{Empirical Results}

\subsection{Robustness to Composed Transformations (RQ1)}
\label{sec:rq1}

\begin{figure}[tbp]
\centering
\includegraphics[width=430.00089pt,height=315.35057pt]{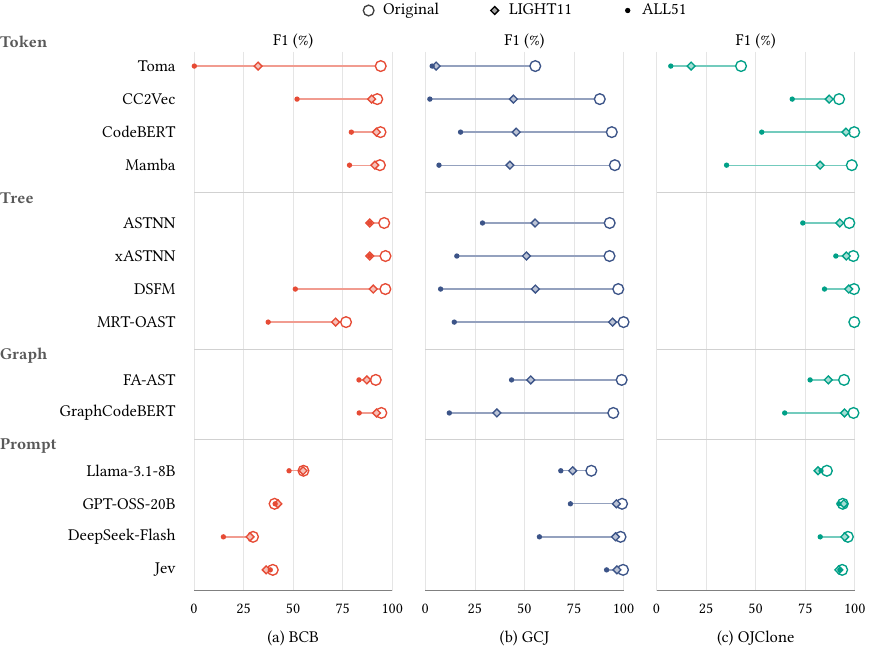}
\caption{Original and transformed F1. Transformed scores are averaged over up to three generation seeds.}
\Description{Three aligned panels compare F1 on BCB, GCJ, and OJClone. Detectors are grouped as token, tree, graph, and prompt models. Each detector has one original marker connected to its available LIGHT11 and ALL51 markers. All markers for a detector share the same horizontal line, with different shapes and sizes distinguishing conditions. MRT-OAST has no transformed evaluation on OJClone.}
\label{fig:rq1_light}
\end{figure}

To examine whether benchmark performance is preserved after code rewriting, we compare each detector on the original code, LIGHT11, and ALL51. Figure~\ref{fig:rq1_light} shows F1 under all three conditions, with transformed scores averaged over the available generation seeds 42, 43, and 44. Table~\ref{tab:rq1_overall} reports seed 42 results, including performance before and after ALL51, transformation coverage, and F1 changes on the full evaluation set and affected pairs.

\textbf{High benchmark performance often fails to hold after composed transformations.}
Figure~\ref{fig:rq1_light} shows that many detectors achieve high F1 on the original code but decline substantially after composed transformations. This gap appears on all three datasets. Table~\ref{tab:rq1_overall} quantifies the pattern. Among token, tree, and graph detectors, 21 of the 25 evaluations starting above 90\% F1 lose at least 10 percentage points under ALL51. CodeBERT falls from 94.06\% to 19.89\% on GCJ, and Mamba from 98.49\% to 35.27\% on OJClone. Despite approaching perfect scores on the original benchmarks, these detectors lose much of their detection performance after behavior-preserving rewrites.

The two transformation configurations in Figure~\ref{fig:rq1_light} show that these losses manifest differently across detectors. On BCB, Toma already loses most of its original performance under LIGHT11, whereas CodeBERT retains more performance under the same configuration. The same detector can also respond very differently to the two configurations. For example, GPT-OSS loses an average of 2.24 F1 points under LIGHT11 on GCJ, compared with 27.18 points under ALL51. Many detectors likewise have LIGHT11 scores between their original and ALL51 scores, particularly on GCJ. The overall decline thus accompanies substantial differences in sensitivity, both across detectors under the same configuration and within a detector across configurations.

\textbf{The overall decline remains stable.}
The overall decline persists when evaluation is restricted to pairs whose code actually changes. Table~\ref{tab:rq1_overall} shows that ALL51 changes 94.14\%--100\% of the pairs in each evaluation, leaving a small proportion unchanged. Excluding these unchanged pairs yields similar performance losses. For example, CodeBERT loses 46.77 F1 points on the full OJClone evaluation set and 49.18 points on affected pairs.

The direction and magnitude of the losses also remain stable across repeated generation. Across token, tree, and graph detectors, all 27 detector and dataset combinations evaluated in full under their native configurations show lower F1 on each of the three independently generated sets of rewrites. The largest standard deviation of the loss is 2.80 points, observed for DSFM on BCB. This variation is small relative to the losses of tens of points reported above, indicating that the overall decline is not an isolated outcome of a particular generation run.

\begin{table}[!t]
\centering
\renewcommand{\arraystretch}{0.90}
\setlength{\aboverulesep}{1.2pt}
\setlength{\belowrulesep}{1.2pt}
\caption{Performance before and after ALL51. $r_{\mathrm{pair}}$ denotes changed-pair coverage. $\Delta F1_{\mathrm{all}}$ and $\Delta F1_{\mathrm{aff}}$ give F1 changes on the full evaluation set and affected pairs, respectively. Dashes indicate missing results due to MRT-OAST's unreleased C frontend.}
\label{tab:rq1_overall}

\fontsize{8.4}{10}\selectfont
\linespread{1}\selectfont
\begin{tabular*}{\linewidth}{l@{\extracolsep{\fill}}llrrrrrrrrr}
\toprule
\multirow{2}{*}{\textbf{Dataset}}
& \multirow{2}{*}{\textbf{Type}}
& \multirow{2}{*}{\textbf{Method}}
& \multicolumn{3}{c}{\textbf{Original}}
& \multicolumn{3}{c}{\textbf{ALL51}}
& \multirow{2}{*}{$r_{\mathrm{pair}}$}
& \multicolumn{2}{c}{\textbf{Change (pp)}} \\
\cmidrule{4-6}
\cmidrule{7-9}
\cmidrule{11-12}
& & & $P$ & $R$ & $F1$
& $P$ & $R$ & $F1$
& & $\Delta F1_{\mathrm{all}}$ & $\Delta F1_{\mathrm{aff}}$ \\
\midrule

\multirow{14}{*}{BCB}
& \multirow{4}{*}{Token}
& Toma & 96.95 & 91.53 & 94.16 & 50.00 & 0.06 & 0.11 & 97.94 & -94.05 & -94.27 \\
& & CC2Vec & 94.47 & 90.42 & 92.40 & 75.34 & 38.15 & 50.65 & 97.20 & -41.76 & -43.02 \\
& & CodeBERT & 93.52 & 94.55 & 94.03 & 89.79 & 69.72 & 78.49 & 97.20 & -15.54 & -15.94 \\
& & Mamba & 94.77 & 92.73 & 93.74 & 87.57 & 70.44 & 78.08 & 97.20 & -15.66 & -16.05 \\
\cmidrule{2-12}
& \multirow{4}{*}{Tree}
& ASTNN & 97.74 & 94.11 & 95.89 & 79.72 & 99.81 & 88.64 & 99.54 & -7.25 & -7.29 \\
& & xASTNN & 99.80 & 93.52 & 96.56 & 79.72 & 99.79 & 88.63 & 99.54 & -7.93 & -7.96 \\
& & DSFM & 96.27 & 96.74 & 96.50 & 81.31 & 39.83 & 53.47 & 97.46 & -43.04 & -44.00 \\
& & MRT-OAST & 83.89 & 70.57 & 76.65 & 68.86 & 26.78 & 38.57 & 98.27 & -38.09 & -38.64 \\
\cmidrule{2-12}
& \multirow{2}{*}{Graph}
& FA-AST & 90.35 & 93.03 & 91.67 & 83.24 & 83.65 & 83.45 & 97.20 & -8.22 & -8.41 \\
& & GraphCodeBERT & 93.46 & 95.52 & 94.48 & 85.56 & 81.43 & 83.44 & 97.20 & -11.04 & -11.30 \\
\cmidrule{2-12}
& \multirow{4}{*}{Prompt}
& Llama & 77.78 & 42.75 & 55.17 & 75.41 & 35.11 & 47.92 & 95.29 & -7.26 & -7.71 \\
& & GPT-OSS-20B & 81.82 & 27.00 & 40.60 & 84.38 & 27.00 & 40.91 & 96.00 & +0.31 & +0.32 \\
& & DeepSeek-Flash & 95.65 & 17.60 & 29.73 & 97.56 & 8.00 & 14.79 & 97.00 & -14.94 & -15.22 \\
& & Jev & 83.87 & 26.00 & 39.69 & 89.09 & 24.50 & 38.43 & 96.00 & -1.26 & -1.27 \\
\midrule

\multirow{14}{*}{GCJ}
& \multirow{4}{*}{Token}
& Toma & 57.32 & 53.70 & 55.45 & 12.80 & 1.98 & 3.43 & 99.64 & -52.02 & -52.35 \\
& & CC2Vec & 90.33 & 85.77 & 87.99 & 6.54 & 1.32 & 2.19 & 99.64 & -85.80 & -86.58 \\
& & CodeBERT & 95.37 & 92.78 & 94.06 & 26.26 & 16.01 & 19.89 & 99.64 & -74.17 & -74.64 \\
& & Mamba & 96.81 & 94.43 & 95.60 & 30.35 & 4.24 & 7.45 & 99.64 & -88.16 & -88.94 \\
\cmidrule{2-12}
& \multirow{4}{*}{Tree}
& ASTNN & 91.35 & 94.70 & 92.99 & 27.94 & 29.63 & 28.76 & 99.64 & -64.23 & -64.55 \\
& & xASTNN & 94.56 & 91.35 & 92.93 & 34.97 & 10.10 & 15.68 & 99.64 & -77.25 & -77.87 \\
& & DSFM & 99.28 & 95.41 & 97.31 & 28.05 & 4.37 & 7.57 & 99.42 & -89.74 & -91.09 \\
& & MRT-OAST
& 100.00 & 100.00 & 100.00
& 27.84 & 9.90 & 14.61
& 100.00 & -85.39 & -85.39 \\
\cmidrule{2-12}
& \multirow{2}{*}{Graph}
& FA-AST
& 99.55 & 98.54 & 99.04
& 44.24 & 42.79 & 43.50
& 100.00 & -55.54 & -55.54 \\
& & GraphCodeBERT & 96.65 & 93.07 & 94.83 & 15.90 & 9.33 & 11.76 & 99.64 & -83.06 & -83.61 \\
\cmidrule{2-12}
& \multirow{4}{*}{Prompt}
& Llama & 72.00 & 100.00 & 83.72 & 62.24 & 75.76 & 68.34 & 98.73 & -15.38 & -15.58 \\
& & GPT-OSS-20B & 100.00 & 98.50 & 99.24 & 86.39 & 63.50 & 73.20 & 98.75 & -26.05 & -26.35 \\
& & DeepSeek-Flash & 100.00 & 97.00 & 98.48 & 100.00 & 40.40 & 57.55 & 99.40 & -40.93 & -41.28 \\
& & Jev & 100.00 & 99.50 & 99.75 & 98.28 & 85.50 & 91.44 & 98.75 & -8.31 & -8.40 \\
\midrule

\multirow{14}{*}{OJClone}
& \multirow{4}{*}{Token}
& Toma & 55.11 & 34.73 & 42.61 & 6.94 & 7.63 & 7.27 & 94.30 & -35.34 & -36.75 \\
& & CC2Vec & 91.31 & 92.81 & 92.06 & 67.79 & 69.01 & 68.40 & 94.30 & -23.66 & -24.90 \\
& & CodeBERT & 99.85 & 99.55 & 99.70 & 52.42 & 53.44 & 52.93 & 94.30 & -46.77 & -49.18 \\
& & Mamba & 99.54 & 97.46 & 98.49 & 55.27 & 25.90 & 35.27 & 94.30 & -63.22 & -67.60 \\
\cmidrule{2-12}
& \multirow{4}{*}{Tree}
& ASTNN & 97.88 & 96.56 & 97.21 & 83.52 & 66.02 & 73.75 & 94.30 & -23.47 & -24.87 \\
& & xASTNN & 99.85 & 98.65 & 99.25 & 92.48 & 88.32 & 90.35 & 94.30 & -8.89 & -9.36 \\
& & DSFM & 99.73 & 99.48 & 99.60 & 88.44 & 81.29 & 84.72 & 94.14 & -14.89 & -15.77 \\
& & MRT-OAST & 99.71 & 99.75 & 99.73 & -- & -- & -- & -- & -- & -- \\
\cmidrule{2-12}
& \multirow{2}{*}{Graph}
& FA-AST & 95.43 & 93.71 & 94.56 & 85.82 & 70.66 & 77.50 & 94.30 & -17.06 & -18.03 \\
& & GraphCodeBERT & 99.85 & 98.80 & 99.32 & 94.03 & 49.55 & 64.90 & 94.30 & -34.42 & -36.75 \\
\cmidrule{2-12}
& \multirow{4}{*}{Prompt}
& Llama & 75.62 & 99.46 & 85.92 & 72.00 & 97.83 & 82.95 & 94.94 & -2.97 & -3.09 \\
& & GPT-OSS-20B & 99.44 & 89.00 & 93.93 & 99.42 & 85.50 & 91.94 & 95.50 & -2.00 & -2.05 \\
& & DeepSeek-Flash & 100.00 & 93.24 & 96.50 & 100.00 & 70.29 & 82.55 & 94.53 & -13.95 & -14.80 \\
& & Jev & 100.00 & 88.00 & 93.62 & 100.00 & 86.00 & 92.47 & 95.50 & -1.14 & -1.18 \\
\bottomrule
\end{tabular*}
\end{table}

\textbf{Original benchmark scores do not reliably rank robustness.}
Detectors with similar original F1 can perform very differently after the same code transformations. On the same OJClone test set, CodeBERT and xASTNN begin at 99.70\% and 99.25\% F1, respectively. Under ALL51, CodeBERT falls to 52.93\%, whereas xASTNN retains 90.35\%. An original gap of less than one percentage point becomes a substantial performance gap after rewriting, with their ranking reversed. On the same GCJ test set, CodeBERT initially exceeds ASTNN at 94.06\% versus 92.99\%, but falls below it after transformation at 19.89\% versus 28.76\%. A detector that scores higher on the original benchmark therefore does not necessarily perform better after rewriting.

Representation category is also insufficient to determine which detector better preserves performance. Consider the graph-based GraphCodeBERT and the token-based CodeBERT. GraphCodeBERT loses less F1 on BCB and OJClone, but more on GCJ. The relative advantage of the same two detectors changes across datasets, so the observed robustness differences cannot be reduced to graph representations outperforming token representations.

\textbf{Prompt-based detectors retain more performance on some datasets.}
On their respective evaluation sets, all four prompt-based detectors lose less F1 on GCJ than the token, tree, and graph detectors. Their largest loss is 40.93 points, whereas the other detectors lose at least 52.02 points. Their F1 after rewriting ranges from 57.55\% to 91.44\%. On OJClone, Llama, GPT-OSS, and Jev also lose less F1 than every evaluated token, tree, and graph detector, while DeepSeek loses 13.95 points, exceeding xASTNN's 8.89-point loss. Overall, prompt-based detectors retain more of their original performance on GCJ and OJClone.

This pattern does not extend uniformly to BCB, where smaller changes also need to be interpreted alongside absolute performance. GPT-OSS, DeepSeek, and Jev start at just 29.73\%--40.60\% F1 on BCB and remain at low scores after rewriting. Although GPT-OSS gains 0.31 F1 points in one generation run, this improvement does not recur in the other two runs, with a mean loss of 1.58 points across all three. In contrast, its substantial GCJ loss persists, averaging 27.18 points. These results show that performance retention among prompt-based detectors varies considerably across datasets, and small changes from low initial scores do not establish effective detection of rewritten clones.

\FloatBarrier
\begingroup
\setlength{\intextsep}{4pt}
\setlength{\columnsep}{0.03\linewidth}
\begin{wraptable}[18]{r}{0.57\linewidth}
\captionsetup{font=small,skip=6pt}
\centering
\caption{Prediction changes under ALL51, reported as percentages of all evaluated pairs. C$\to$W denotes correct predictions becoming wrong after transformation, and W$\to$C the reverse. Dashes indicate missing results due to MRT-OAST's unreleased C frontend.}
\label{tab:rq1_transitions}
\small
\linespread{1}\selectfont
\setlength{\tabcolsep}{2pt}
\renewcommand{\arraystretch}{1.0}
\begin{tabular*}{\linewidth}{@{\hspace{\tabcolsep}\extracolsep{\fill}}lrrrrrr}
\toprule
\multirow{2}{*}{Method} & \multicolumn{2}{c}{BCB} & \multicolumn{2}{c}{GCJ} & \multicolumn{2}{c}{OJClone} \\
\cmidrule{2-3}\cmidrule{4-5}\cmidrule{6-7}
 & C$\to$W & W$\to$C & C$\to$W & W$\to$C & C$\to$W & W$\to$C \\
\midrule
Toma & 8.32 & 0.26 & 12.80 & 7.87 & 8.88 & 2.12 \\
CC2Vec & 8.73 & 0.59 & 20.18 & 1.81 & 3.50 & 0.31 \\
CodeBERT & 4.29 & 0.71 & 24.01 & 1.13 & 6.33 & 0.02 \\
Mamba & 4.24 & 0.52 & 19.51 & 0.61 & 6.19 & 0.04 \\
\cmidrule{1-7}
ASTNN & 18.61 & 4.68 & 27.47 & 1.60 & 2.87 & 0.10 \\
xASTNN & 20.18 & 5.13 & 19.26 & 0.76 & 1.18 & 0.02 \\
DSFM & 9.86 & 0.50 & 21.38 & 0.17 & 1.91 & 0.01 \\
MRT-OAST & 17.82 & 3.70 & 22.62 & 0.00 & -- & -- \\
\cmidrule{1-7}
FA-AST & 3.26 & 1.02 & 21.43 & 0.09 & 2.23 & 0.21 \\
GraphCodeBERT & 3.60 & 0.70 & 25.97 & 0.62 & 3.50 & 0.01 \\
\cmidrule{1-7}
Llama-3.1-8B & 9.76 & 6.73 & 25.13 & 9.39 & 7.30 & 3.37 \\
GPT-OSS-20B & 3.50 & 4.00 & 22.50 & 0.00 & 3.00 & 1.25 \\
DeepSeek-Flash & 5.20 & 0.70 & 28.30 & 0.00 & 11.74 & 0.40 \\
Jev & 1.50 & 1.75 & 7.75 & 0.00 & 1.00 & 0.00 \\
\bottomrule
\end{tabular*}
\end{wraptable}

\textbf{Prediction changes predominantly introduce new errors.}
Code rewriting turns previously correct judgments into errors more often than it corrects existing mistakes. Table~\ref{tab:rq1_transitions} reports the proportions of correct-to-wrong and wrong-to-correct transitions. On GCJ and OJClone, every detector with available results has more correct-to-wrong transitions than the reverse. For CodeBERT on GCJ, 24.01\% of evaluated pairs change from correct to wrong, whereas just 1.13\% change from wrong to correct.

Individual predictions need not remain stable even when aggregate scores change little. On BCB, GPT-OSS gains just 0.31 F1 points, yet 3.50\% of pairs change from correct to wrong and 4.00\% change in the opposite direction. Beneath the nearly unchanged score, the set of correctly classified pairs has changed. Evaluating performance after rewriting therefore requires examining not just aggregate scores but also whether a detector preserves its correct judgments on the same code pairs.

\par\endgroup

\begin{resultbox}
\textbf{Finding 1.}
High performance on the original benchmark does not guarantee that a detector remains effective after behavior-preserving code rewriting. Of the 25 token, tree, and graph detector and dataset combinations starting above 90\% F1, 21 lose at least 10 percentage points under ALL51. Detector rankings also reverse on shared test sets, showing that original performance advantages need not persist after rewriting. Moreover, nearly unchanged aggregate scores can conceal the loss of previously correct predictions.
\end{resultbox}

\subsection{Sensitivity to Transformation Type and Intensity (RQ2)}
\label{sec:rq2}

\begin{figure}[p]
\centering
\raisebox{-254.69618pt}[259.30869pt][254.69618pt]{%
  \includegraphics[width=428.7514pt,height=514.00487pt]{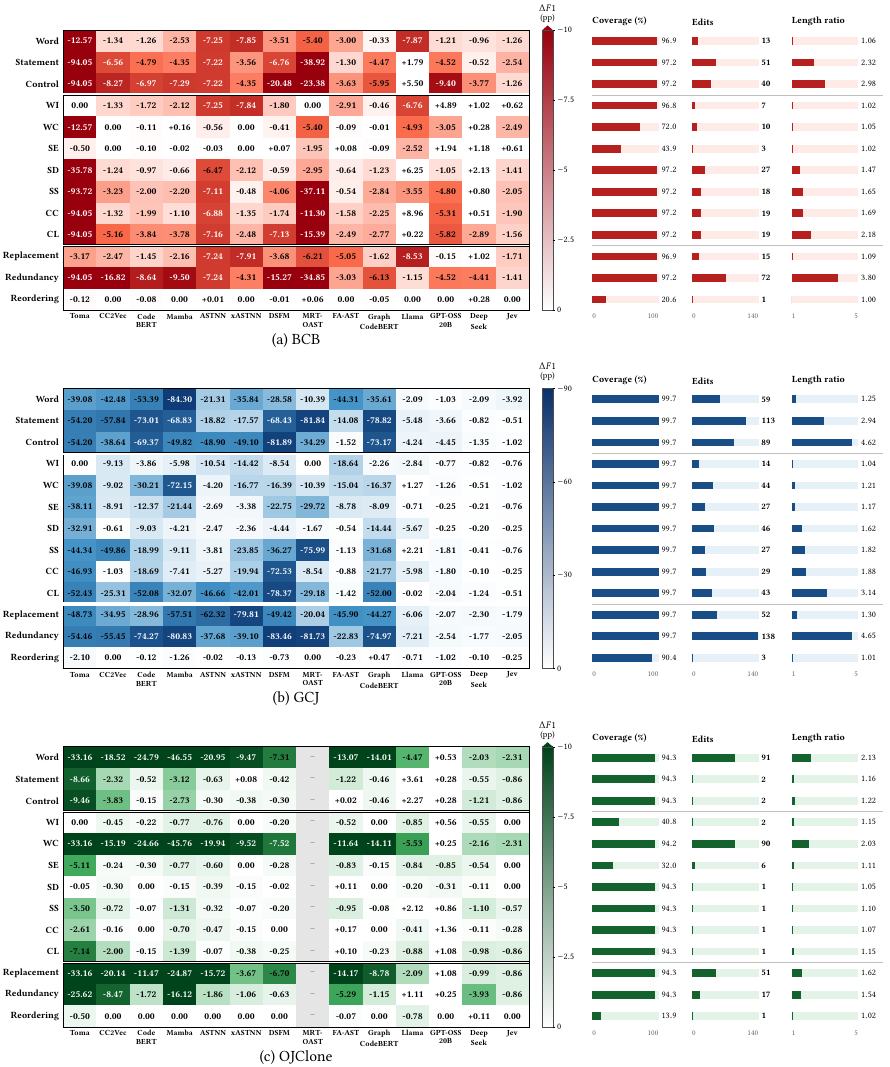}}
\captionsetup{font=footnotesize,skip=4pt}
\par\begin{minipage}{\linewidth}
\caption{Grouped SCT effects on (a) BCB, (b) GCJ, and (c) OJClone. Bars on the right report the percentage of designated programs changed and, over changed programs, median successful edits and median length ratios (transformed/original UTF-8 bytes). Dashes mark missing results due to MRT-OAST's unreleased C frontend. Color ranges differ across datasets, and values outside each range use the corresponding endpoint color.}
\label{fig:rq2_grouped}
\end{minipage}
\end{figure}

The overall losses in RQ1 leave open which code changes detectors are sensitive to and how their effects develop with further rewriting. We first compare the 13 transformation groups, then examine repeated application of eight operators supporting continued insertion or expansion. Figure~\ref{fig:rq2_grouped} reports group effects alongside coverage, edit counts, and code length ratios, allowing performance changes to be interpreted in relation to the rewriting each group achieves.

\textbf{Shared sensitivities emerge across transformation groups.}
Replacement and redundancy produce larger absolute F1 changes than reordering in all 29 available evaluations of token, tree, and graph detectors. Reordering changes F1 by less than one point in 27 evaluations, with a largest decrease of 2.10 points. On GCJ, for example, CodeBERT loses 74.27 points under redundancy and 28.96 under replacement, compared with 0.12 under reordering.
On GCJ, reordering affects 90.36\% of programs, showing that its small impact on F1 cannot be explained simply by low transformation coverage.

Across code levels, the Word composition lowers F1 in all 29 evaluations. On OJClone, it causes the largest loss among the three levels for each of the nine available detectors, while constants cause the largest loss among the seven targets. Word-level sensitivity thus extends across token, tree, and graph representations.

\textbf{Sensitive targets depend on the detector and the corpus.}
The most sensitive transformation target differs among detectors. On GCJ, constants cause the largest target-level loss for Mamba, ordinary statements for CC2Vec, and loops for CodeBERT, although all three use token representations. A shared representation category therefore does not imply a shared sensitive target.

This variation also extends across datasets. Loop transformations cause the largest target-level losses for five token, tree, or graph detectors on GCJ, whereas constants dominate for all nine available detectors on OJClone. On OJClone, constant transformations reduce CodeBERT, ASTNN, and GraphCodeBERT by 24.66, 19.94, and 14.11 points, respectively, revealing a shared weakness across representation categories. Sensitive targets therefore need to be examined within each dataset rather than inferred from representation type alone.

\begin{figure}[htbp]
\centering
\includegraphics[width=\linewidth]{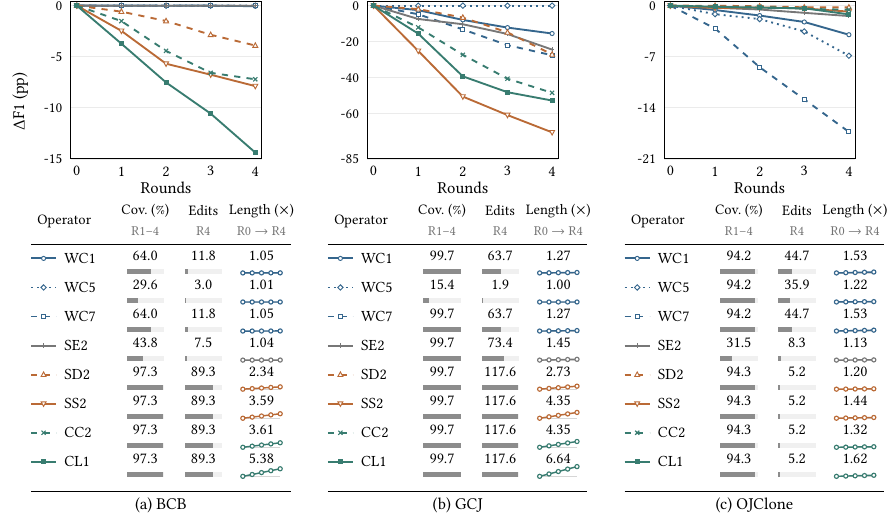}
\caption{F1 response and realized transformation intensity. Curves show full-set $\Delta$F1 medians across ten token, tree, and graph detectors on BCB/GCJ and nine on OJClone, excluding MRT-OAST because its C frontend is unreleased. F1 scales vary by dataset. Coverage is constant over R1--4. Edits and byte-length ratios are detector medians of program means, including unchanged programs. Length labels show R4 ratios. Lower panels share scales across datasets, with coverage 0--100\%, edits 0--120, and length 1--7$\times$.}
\label{fig:rq2_intensity}
\end{figure}

\textbf{Repeated application produces different response trajectories.}
To examine how performance changes as rewriting accumulates, we apply each of the eight operators supporting continued insertion or expansion in Table~\ref{tab:operators} over four rounds. Figure~\ref{fig:rq2_intensity} traces the resulting changes from the original code at round 0. The median trajectories show both accelerating losses and diminishing additional losses. On GCJ, the median loss under declaration insertion (SD2) grows from 1.93 points at round 1 to 27.05 at round 4, with larger increments in later rounds. Computation insertion (SS2) instead adds about 25 points of loss in each of the first two rounds, followed by roughly ten points in each subsequent round, reaching 70.34 points at round 4. On BCB, the insertion operators show accumulating median losses, while the median changes under WC1, WC5, WC7, and SE2 remain close to zero.

\begin{figure}[htbp]
\centering
\includegraphics[width=430.00089pt,height=280.53268pt]{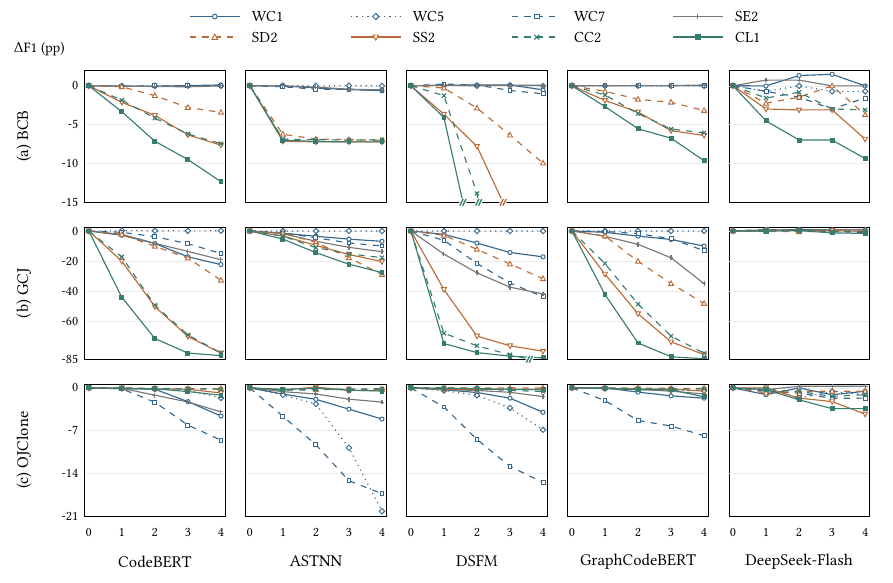}
\caption{F1 change for five representative detectors after one to four applications per original site of the eight operators supporting continued insertion or expansion. Rows give datasets and columns give detectors, with a shared vertical scale within each dataset. Double slashes mark curves crossing the axis limits.}
\label{fig:rq2_representatives}
\end{figure}

Individual detectors also differ under the same repeated operator. Figure~\ref{fig:rq2_representatives} shows trajectories for five detectors. On BCB, ASTNN incurs most of its SS2 loss in the first round and then remains nearly unchanged. CodeBERT and GraphCodeBERT continue to lose F1, while DSFM's loss grows more sharply in later rounds. On GCJ, DeepSeek remains close to its original F1 over four rounds. Repetition can therefore produce an early drop followed by a plateau, sustained degradation, or little change, depending on both the detector and the operator.

\textbf{Application rounds do not directly measure rewriting amounts, and rewriting amounts alone do not determine performance losses.}
Application rounds count repeated applications of an operator, while the number of edits in each round depends on the number of eligible locations in a program. Equal rounds therefore need not correspond to equal amounts of rewriting. To examine the relationship between rewriting intensity and performance losses, we further compare realized edit counts and code growth.

More edits do not necessarily produce larger performance losses. On OJClone, WC1 and WC5 both reach about 94.2\% coverage. At round 4, WC1 accumulates about 44.7 edits, exceeding WC5's 35.9, yet its median F1 loss is 3.99 points, below WC5's 6.87 points. Similar rewriting amounts can also produce substantially different losses. On GCJ, computation insertion (SS2) and branch insertion (CC2) both reach about 99.7\% coverage and accumulate about 117.6 edits by round 4, with code length increasing to approximately 4.35 times the original. Their median F1 losses nevertheless differ at 70.34 and 48.39 points, respectively. These comparisons show that edit counts and code growth alone do not determine performance losses, while the specific rewritten content also shapes detector responses.

\begin{resultbox}
\textbf{Finding 2.}
Detector sensitivity to behavior-preserving rewriting depends on the transformation type and how it is applied. Replacement and redundancy produce larger absolute F1 changes than reordering in all 29 evaluations of token, tree, and graph detectors, while the most sensitive target varies across detectors and datasets. Repeated application can produce sustained degradation, an initial drop followed by a plateau, or little change. More edits do not necessarily cause larger losses, and different rewrites can produce substantially different performance declines even when coverage, edit counts, and code growth are similar.
\end{resultbox}

\FloatBarrier
\subsection{Alternative Constructions within Transformation Families (RQ3)}
\label{sec:rq3}

\begin{figure}[htbp]
\centering
\includegraphics[width=430.00089pt,height=290.48065pt]{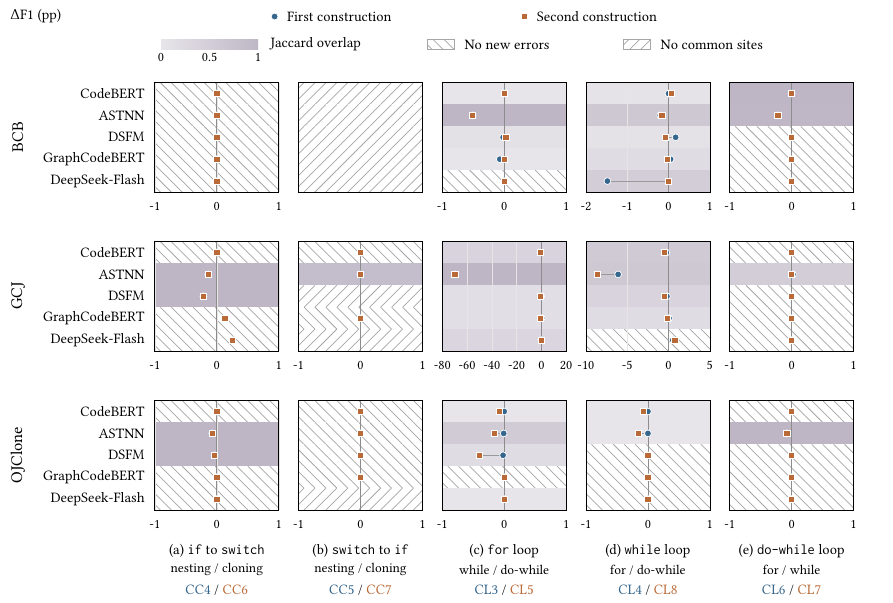}
\caption{Five pairs of alternative constructions within transformation families, applied once at common sites with matching edit counts. Rows give datasets and columns give construction comparisons. Circles and squares show the F1 changes for the first and second constructions over each detector's evaluation set. Muted purple shading gives the Jaccard overlap of newly introduced error sets. Diagonal lines sloping down to the right indicate that neither construction introduces new errors, while lines sloping down to the left indicate no common applicable sites.}
\label{fig:rq3_construction}
\end{figure}

The same behavior-preserving transformation can be implemented through different code constructions. Even at identical locations and edit counts, the resulting syntactic structures may differ. Summarizing performance changes by transformation family can therefore conceal detector sensitivity to specific constructions. We compare five pairs of conditional and loop conversions, applying each alternative once at common applicable locations in the original programs with matching edit counts. Figure~\ref{fig:rq3_construction} reports both the F1 change over each detector's evaluation set and the overlap of newly introduced errors. A newly introduced error is a pair classified correctly before rewriting and incorrectly afterward. The Jaccard overlap measures the intersection of the two error sets divided by their union.

\textbf{Construction choices change performance losses, but their effects vary across families and detectors.}
On GCJ, converting a \texttt{while} loop to a \texttt{for} loop or a \texttt{do-while} loop reduces ASTNN's F1 by 6.11 and 8.64 points, respectively. Both constructions act at the same program locations with matching edit counts, yet produce different performance losses. Thus, even with locations and edit counts fixed, the choice of construction affects the evaluation outcome.

Not every construction choice produces a different outcome. For the two \texttt{if}-to-\texttt{switch} constructions, each of the four token, tree, and graph detectors has identical F1 scores and newly introduced error sets under both alternatives in all three datasets. Identical outcomes need not indicate robustness. On GCJ, both conversions from a \texttt{for} loop reduce ASTNN's F1 by 69.79 points and produce identical newly introduced error sets. Construction dependence thus varies with the transformation family and the detector, rather than following directly from the size of the loss.

\textbf{Similar aggregate losses can conceal different newly introduced errors.}
On GCJ, converting a \texttt{for} loop to a \texttt{while} loop or a \texttt{do-while} loop reduces CodeBERT's F1 by 0.90 and 1.06 points. Despite these similar aggregate losses, the newly introduced error sets have a Jaccard overlap of just 0.38, indicating substantial differences in the failures exposed by the two constructions. For the same conversion pair and dataset, ASTNN has an overlap of 1.00. F1 changes therefore measure the magnitude of performance loss, while error overlap further reveals whether alternative constructions cause the same pairs to be misclassified.

\begin{resultbox}
\textbf{Finding 3.}
Within a transformation family, alternative constructions can produce different performance losses and expose different errors at matched locations and edit counts. Similar aggregate losses do not imply that the same pairs fail, while some alternatives produce identical outcomes. Evaluating multiple constructions reveals sensitivities that a single construction may leave unobserved.
\end{resultbox}

\FloatBarrier
\subsection{Discrimination and Failure Analysis (RQ4)}
\label{sec:rq4}

F1 losses may reflect missed clones, accepted non-clones, or a mismatch between scores and the fixed threshold. We examine class-specific errors, score distributions, and threshold adjustment under ALL51, followed by concrete error cases and the effects of evaluation conditions.

\subsubsection{Discrimination and Threshold Adjustment}
\label{sec:rq4_discrimination}

The key distinction is whether rewriting weakens discrimination between clones and non-clones or shifts the operating point of the fixed threshold.

\begin{figure}[ht]
\centering
\includegraphics[width=430.00089pt,height=209.90211pt]{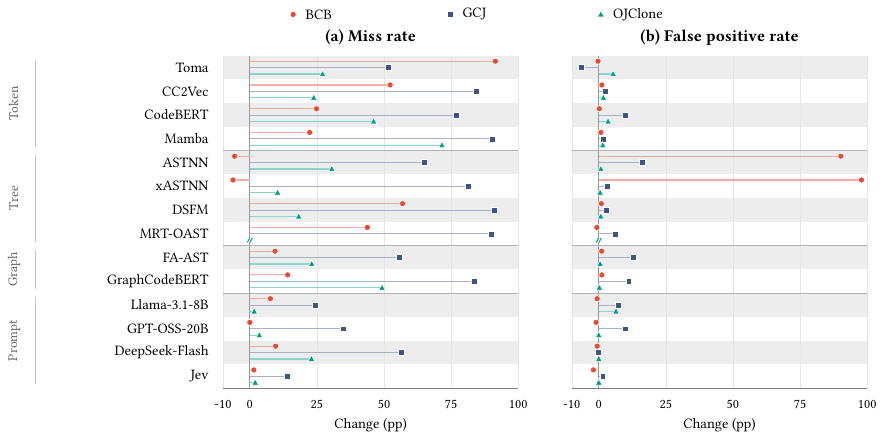}
\caption{Changes in (a) miss rate and (b) false positive rate under ALL51 at seed 42. Points show transformed minus original error rates in percentage points on a shared scale. Negative values indicate decreases. Double slashes denote unavailable evaluation.}
\label{fig:rq4_errors}
\end{figure}

\textbf{Error direction.}
Figure~\ref{fig:rq4_errors} separates changes in miss rate and false positive rate. On GCJ, DSFM's recall falls from 95.41\% to 4.37\%, while its false positive rate rises from 0.18\% to 2.96\%, primarily reflecting failure to recognize rewritten clones. In contrast, ASTNN and xASTNN retain recall near 99.8\% on BCB, but their false positive rates rise to 98.62\% and 98.57\%, indicating that they reject almost no non-clone pairs. Among prompt-based detectors, DeepSeek and Jev retain precision above 98\% on GCJ while losing recall, whereas GPT-OSS also increases its false positive rate from 0\% to 10\%. These results distinguish degradation dominated by missed clones from degradation involving increased acceptance of non-clones.

\begin{figure}[htbp]
\centering
\includegraphics[width=430.01765pt,height=302.52610pt]{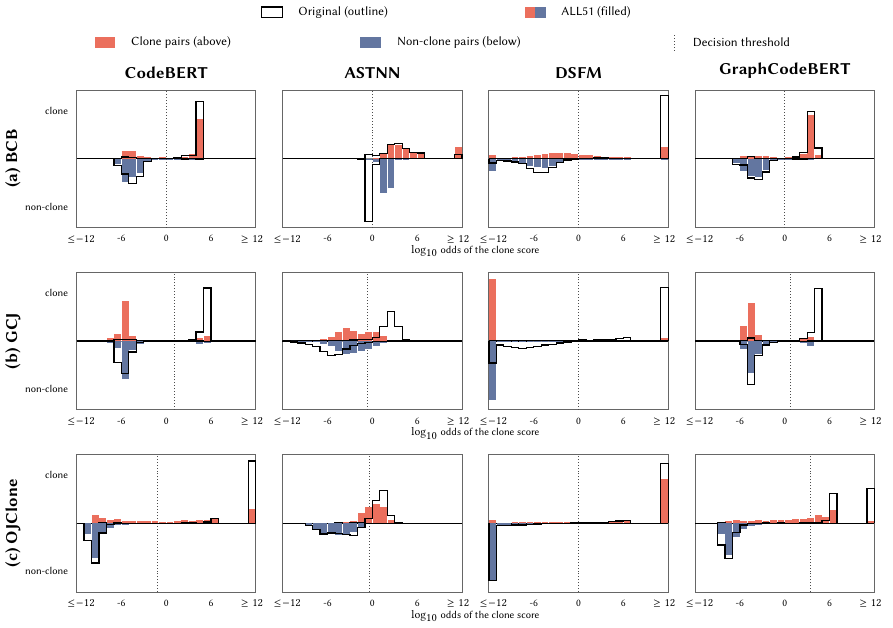}
\caption{Clone scores before and after ALL51 at seed 42. Clone pairs appear above the axis and non-clone pairs below it, as the share of each class in every bin of $\log_{10}$ odds. Scores beyond $\pm12$ are included in the end bins. Black outlines give original scores, colored bars give transformed scores, and dotted lines mark the fixed thresholds.}
\label{fig:rq4_scores}
\end{figure}

\textbf{Score separability.}
Figure~\ref{fig:rq4_scores} shows how the distributions of clone and non-clone scores change, while Table~\ref{tab:rq4_auc} quantifies their separation using ROC AUC. On GCJ, the transformed clone and non-clone scores of CodeBERT, ASTNN, DSFM, and GraphCodeBERT concentrate in similar regions, with both classes heavily concentrated at the low end for DSFM. Increased misses thus accompany weaker separation between clone and non-clone scores. The GCJ results in Table~\ref{tab:rq4_auc} further support this observation. Seven of the eight token, tree, and graph detectors with available scores have AUC falling from an original range of 95.88\%--100\% to 36.56\%--57.27\%, indicating ranking near chance or in the reverse direction. However, FA-AST and Jev retain transformed AUC of 75.14\% and 98.55\% on GCJ, showing that the deterioration in separability varies across detectors.

A decline in F1 can also coexist with strong score separability. On OJClone, Figure~\ref{fig:rq4_scores} shows ASTNN's transformed clone scores moving toward the fixed threshold, with some falling below it, while remaining generally higher than non-clone scores. Its F1 falls from 97.21\% to 73.75\%, yet its AUC remains at 98.14\%. Thus, although rewriting increases classification errors at the fixed threshold, the ordering of the two classes remains largely preserved.

\textbf{Threshold adjustment.}
Table~\ref{tab:rq4_auc} reports the maximum F1 over thresholds on the transformed test set. This oracle uses test labels and gives a diagnostic upper bound for threshold adjustment in the existing score direction, rather than a deployable recalibration procedure. Gain measures the increase from fixed-threshold F1 to oracle F1. For the seven GCJ detectors with AUC near or below chance, threshold adjustment improves F1 by 4.63--26.99 percentage points, but oracle F1 remains between 32.67\% and 35.04\%. FA-AST reaches 47.18\% oracle F1, well below its original 99.04\%. On OJClone, ASTNN's oracle F1 is 76.91\%, despite its retained high AUC. These comparisons show that threshold adjustment does not restore the original performance of the token, tree, and graph detectors with available scores.

Jev illustrates a different balance between score separability and the fixed threshold. On GCJ, threshold adjustment raises its transformed F1 from 91.44\% to 95.50\%, recovering part of the decline from 99.75\%. On BCB, its transformed AUC is 83.34\%, while F1 rises from 38.43\% at the fixed threshold to 78.42\% at the oracle threshold. Here the fixed decision boundary leaves substantial usable separation in the scores unexploited.

\FloatBarrier
\begin{table}[htbp]
\centering
\caption{ROC AUC and F1 under ALL51 at seed 42. Subscripts 0 and $T$ denote original and transformed code. Oracle F1 maximizes F1 over thresholds using transformed test labels. Gain is Oracle F1$_T$ minus F1$_T$ in percentage points, calculated before rounding.}
\label{tab:rq4_auc}
\begingroup
\small
\sisetup{round-mode=places,round-precision=2}
\setlength{\tabcolsep}{4pt}
\renewcommand{\arraystretch}{1.06}
\begin{tabular*}{\linewidth}{@{\hspace{\tabcolsep}\extracolsep{\fill}}ccl*{4}{S[table-format=3.2]}*{2}{S[table-format=2.2]}}
\toprule
\multirow{2}{*}{\textbf{Dataset}} & \multirow{2}{*}{\textbf{Type}} & \multicolumn{1}{c}{\multirow{2}{*}{\textbf{Method}}} & \multicolumn{2}{c}{\textbf{Ranking}} & \multicolumn{2}{c}{\textbf{Fixed threshold}} & \multicolumn{2}{c}{\textbf{Threshold diagnostics}} \\
\cmidrule[0.3pt](lr){4-5}\cmidrule[0.3pt](lr){6-7}\cmidrule[0.3pt](lr){8-9}
 & & & {\textbf{AUC$_0$}} & {\textbf{AUC$_T$}} & {\textbf{F1$_0$}} & {\textbf{F1$_T$}} & {\textbf{Oracle F1$_T$}} & {\textbf{Gain}} \\
\midrule[0.6pt]
\multirow{10}{*}[-6pt]{BCB} & \multirow{3}{*}{Token} & Toma & 99.35 & 49.42 & 94.16 & 0.11 & 16.67 & 16.56 \\
 &  & CodeBERT & 96.82 & 86.13 & 94.03 & 78.49 & 79.00 & 0.51 \\
 &  & Mamba & 98.17 & 92.57 & 93.74 & 78.08 & 78.69 & 0.61 \\
\cmidrule[0.3pt]{2-9}
 & \multirow{4}{*}{Tree} & ASTNN & 96.69 & 91.12 & 95.89 & 88.64 & 90.59 & 1.95 \\
 &  & xASTNN & 96.82 & 86.77 & 96.56 & 88.63 & 88.64 & 0.01 \\
 &  & DSFM & 98.84 & 80.51 & 96.50 & 53.47 & 59.82 & 6.36 \\
 &  & MRT-OAST & 86.88 & 59.30 & 76.65 & 38.57 & 50.00 & 11.43 \\
\cmidrule[0.3pt]{2-9}
 & \multirow{2}{*}{Graph} & FA-AST & 98.77 & 95.61 & 91.67 & 83.45 & 83.80 & 0.36 \\
 &  & GraphCodeBERT & 97.34 & 91.47 & 94.48 & 83.44 & 84.61 & 1.17 \\
\cmidrule[0.3pt]{2-9}
 & Prompt & Jev & 82.53 & 83.34 & 39.69 & 38.43 & 78.42 & 39.99 \\
\midrule[0.3pt]
\multirow{9}{*}[-6pt]{GCJ} & \multirow{2}{*}{Token} & CodeBERT & 96.80 & 46.56 & 94.06 & 19.89 & 32.69 & 12.80 \\
 &  & Mamba & 99.43 & 52.02 & 95.60 & 7.45 & 32.70 & 25.25 \\
\cmidrule[0.3pt]{2-9}
 & \multirow{4}{*}{Tree} & ASTNN & 99.09 & 56.61 & 92.99 & 28.76 & 33.39 & 4.63 \\
 &  & xASTNN & 98.90 & 57.27 & 92.93 & 15.68 & 35.04 & 19.36 \\
 &  & DSFM & 99.62 & 46.86 & 97.31 & 7.57 & 34.55 & 26.99 \\
 &  & MRT-OAST & 100.00 & 45.67 & 100.00 & 14.61 & 32.69 & 18.08 \\
\cmidrule[0.3pt]{2-9}
 & \multirow{2}{*}{Graph} & FA-AST & 99.66 & 75.14 & 99.04 & 43.50 & 47.18 & 3.68 \\
 &  & GraphCodeBERT & 95.88 & 36.56 & 94.83 & 11.76 & 32.67 & 20.90 \\
\cmidrule[0.3pt]{2-9}
 & Prompt & Jev & 99.50 & 98.55 & 99.75 & 91.44 & 95.50 & 4.06 \\
\midrule[0.3pt]
\multirow{9}{*}[-6pt]{OJClone} & \multirow{3}{*}{Token} & Toma & 80.04 & 50.03 & 42.61 & 7.27 & 12.52 & 5.25 \\
 &  & CodeBERT & 99.96 & 89.42 & 99.70 & 52.93 & 54.24 & 1.31 \\
 &  & Mamba & 99.94 & 77.32 & 98.49 & 35.27 & 37.39 & 2.12 \\
\cmidrule[0.3pt]{2-9}
 & \multirow{3}{*}{Tree} & ASTNN & 99.87 & 98.14 & 97.21 & 73.75 & 76.91 & 3.17 \\
 &  & xASTNN & 99.90 & 99.51 & 99.25 & 90.35 & 90.60 & 0.25 \\
 &  & DSFM & 99.98 & 96.78 & 99.60 & 84.72 & 84.94 & 0.23 \\
\cmidrule[0.3pt]{2-9}
 & \multirow{2}{*}{Graph} & FA-AST & 99.57 & 98.36 & 94.56 & 77.50 & 78.97 & 1.47 \\
 &  & GraphCodeBERT & 99.89 & 96.08 & 99.32 & 64.90 & 74.35 & 9.44 \\
\cmidrule[0.3pt]{2-9}
 & Prompt & Jev & 99.87 & 99.84 & 93.62 & 92.47 & 98.26 & 5.79 \\
\bottomrule
\end{tabular*}
\endgroup
\end{table}

\subsubsection{Error Case Analysis}
\label{sec:rq4_error_cases}

\begin{figure}[htbp]
\centering
\raisebox{-239.3539pt}[5.4343pt][239.3539pt]{%
  \includegraphics[width=430.00827pt,height=244.78820pt]{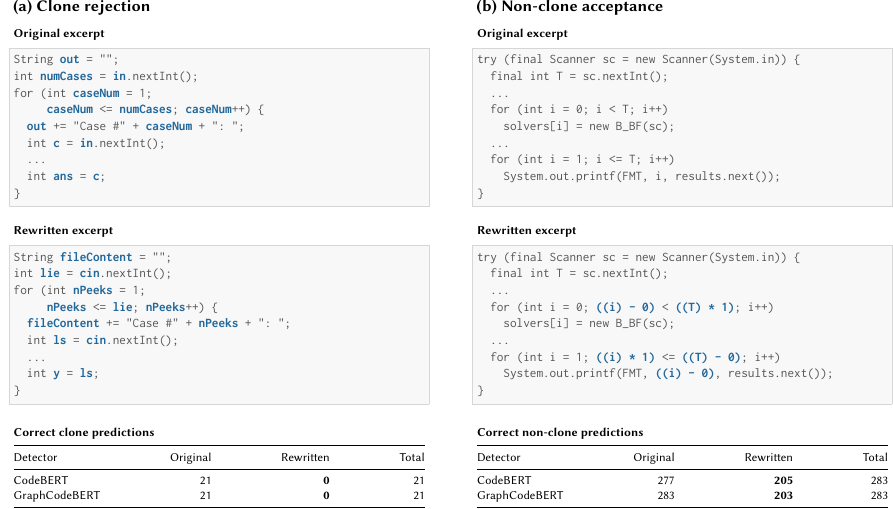}}
\caption{Behavior-preserving edits and prediction changes on GCJ at round 1 and seed 42. In (a), WI1 renames 15 identifiers, making both detectors reject all 21 previously detected clone pairs. In (b), SE2 inserts five neutral arithmetic wrappers, increasing non-clone errors. Blue bold text highlights the changes. Edit counts cover the full programs, while code is excerpted. Prediction counts refer to pairs containing each program. FMT denotes the output format string.}
\label{fig:rq4_cases}
\end{figure}

The aggregate results identify changes in error direction and score separability. Figure~\ref{fig:rq4_cases} examines two concrete examples of how behavior-preserving edits alter clone decisions, covering both rejection of clones and acceptance of non-clones.

\textbf{Clone rejection after identifier renaming.}
In Figure~\ref{fig:rq4_cases}(a), WI1 consistently renames identifiers and their references, such as \texttt{numCases} to \texttt{lie} and \texttt{caseNum} to \texttt{nPeeks}. The renamed variables retain their bindings, while the operations, control flow, and output strings remain unchanged. CodeBERT and GraphCodeBERT nevertheless reject all previously recognized clone pairs containing this program. This case shows that changes to identifier spelling can disrupt clone recognition even when the program's computation is preserved.

\textbf{Non-clone acceptance after expression rewriting.}
In Figure~\ref{fig:rq4_cases}(b), SE2 wraps integer expressions in neutral arithmetic, replacing \texttt{i} with \texttt{i - 0} or \texttt{i * 1} and \texttt{T} with \texttt{T * 1} or \texttt{T - 0}. These expressions retain their values, so the loop conditions and printed arguments are unchanged. The rewritten program therefore retains its non-clone relationships with programs solving different problems. Both detectors nevertheless accept more non-clone pairs. Together, the cases show that behavior-preserving edits can both remove correct clone decisions and introduce false similarities, complementing the class-specific error patterns in Figure~\ref{fig:rq4_errors}.

\FloatBarrier
\subsubsection{Effects of Evaluation Conditions}
\label{sec:rq4_evaluation_conditions}

The preceding analyses characterize discrimination under frozen evaluation configurations. To examine how these results depend on input processing, we compare prediction flips across estimated input limits and assess whether the performance gaps persist when batch size changes.

\Needspace{14\baselineskip}
\begingroup
\setlength{\columnsep}{0.035\linewidth}
\setlength{\intextsep}{6pt}
\begin{wraptable}{r}{0.60\linewidth}
\centering
\captionsetup{font=footnotesize,skip=5pt}
\caption{Estimated input-limit crossings and prediction flips after rewriting. Newly over limit gives the proportion of inputs pushed over the limit. Within limit and Over limit give matched flip rates for inputs that remain within the limit and those newly over it, respectively. Gap is Over limit minus Within limit.}
\label{tab:rq4_truncation}
\begingroup
\footnotesize
\linespread{1}\selectfont
\sisetup{round-mode=none}
\setlength{\tabcolsep}{4pt}
\renewcommand{\arraystretch}{1.08}
\begin{tabular}{clS[table-format=2.1]S[table-format=2.1]S[table-format=2.1]S[table-format=2.1]}
\toprule
\multirow{2}{*}[-2pt]{\textbf{Dataset}} & \multicolumn{1}{c}{\multirow{2}{*}[-2pt]{\textbf{Method}}} & \multicolumn{1}{c}{\multirow{2}{*}[-2pt]{\makecell{\textbf{Newly over}\\\textbf{limit (\%)}}}} & \multicolumn{3}{c}{\textbf{Prediction flips}} \\
\cmidrule[0.3pt]{4-6}
 & & & {\textbf{Within limit}} & {\textbf{Over limit}} & {\textbf{Gap (pp)}} \\
\midrule[0.6pt]
\multirow{3}{*}{BCB} & CodeBERT & 37.2 & 0.4 & 6.4 & 5.9 \\
 & GraphCodeBERT & 37.2 & 0.0 & 5.5 & 5.5 \\
 & Mamba & 23.6 & 1.8 & 16.1 & 14.3 \\
\midrule[0.3pt]
\multirow{3}{*}{GCJ} & CodeBERT & 52.0 & 8.8 & 15.8 & 6.9 \\
 & GraphCodeBERT & 52.1 & 8.1 & 13.9 & 5.8 \\
 & Mamba & 45.9 & 27.0 & 35.8 & 8.8 \\
\midrule[0.3pt]
\multirow{3}{*}{OJClone} & CodeBERT & 16.3 & 8.5 & 14.3 & 5.9 \\
 & GraphCodeBERT & 16.1 & 6.6 & 17.6 & 11.0 \\
 & Mamba & 3.4 & 65.0 & 70.6 & 5.6 \\
\bottomrule
\end{tabular}
\endgroup
\end{wraptable}
\textbf{Input length.}
We examine CodeBERT, GraphCodeBERT, and Mamba because their input pipelines truncate code to fit finite token budgets. Code growth can therefore change which parts of a program reach the detector. Table~\ref{tab:rq4_truncation} compares prediction flips among originally correct clone pairs with changed code across 46 conditions at seed 42, covering ALL51, 13 group compositions, and rounds 1--4 of the eight operators supporting continued insertion or expansion. We compare inputs under the same transformation condition and with similar relative code growth, balancing the contributions of the two groups. The comparison uses compact CodeBERT token counts, with estimated limits of 398 tokens per rewritten fragment for CodeBERT and GraphCodeBERT and 1,020 tokens per pair for Mamba. Under these matched conditions, longer inputs are associated with more prediction errors. Across all nine comparisons, inputs crossing the estimated limit flip more frequently, while inputs remaining within it also exhibit prediction flips. Sensitivity to rewriting therefore extends beyond inputs crossing the estimated length limit to those remaining within it.
\par\endgroup

\Needspace{12\baselineskip}
\begingroup
\setlength{\columnsep}{0.035\linewidth}
\setlength{\intextsep}{6pt}
\begin{wraptable}{r}{0.60\linewidth}
\captionsetup{font=footnotesize,skip=5pt}
\centering
\caption{Batch size controls for ASTNN and xASTNN. BS denotes batch size. $\Delta$F1 is ALL51 minus Original F1 in percentage points.}
\label{tab:rq4_batch_context}
\begingroup
\small
\linespread{1}\selectfont
\sisetup{round-mode=none}
\setlength{\tabcolsep}{5pt}
\renewcommand{\arraystretch}{1.08}
\begin{tabular}{cl*{4}{S[table-format=2.2]}*{2}{S[table-format=-2.2]}}
\toprule
\multirow{2}{*}[-2pt]{\textbf{Dataset}} & \multicolumn{1}{c}{\multirow{2}{*}[-2pt]{\textbf{Method}}} & \multicolumn{2}{c}{\textbf{Original F1}} & \multicolumn{2}{c}{\textbf{ALL51 F1}} & \multicolumn{2}{c}{\textbf{$\Delta$F1 (pp)}} \\
\cmidrule[0.3pt]{3-4}\cmidrule[0.3pt]{5-6}\cmidrule[0.3pt]{7-8}
 & & {\textbf{BS32}} & {\textbf{BS1}} & {\textbf{BS32}} & {\textbf{BS1}} & {\textbf{BS32}} & {\textbf{BS1}} \\
\midrule[0.6pt]
\multirow{2}{*}{BCB} & ASTNN & 95.89 & 88.83 & 88.64 & 88.59 & -7.25 & -0.24 \\
 & xASTNN & 96.56 & 89.35 & 88.63 & 88.58 & -7.93 & -0.77 \\
\midrule[0.3pt]
\multirow{2}{*}{GCJ} & ASTNN & 92.99 & 93.02 & 28.76 & 26.85 & -64.23 & -66.18 \\
 & xASTNN & 92.93 & 93.02 & 15.68 & 10.70 & -77.25 & -82.32 \\
\midrule[0.3pt]
\multirow{2}{*}{OJClone} & ASTNN & 97.21 & 93.81 & 73.75 & 66.02 & -23.47 & -27.79 \\
 & xASTNN & 99.25 & 99.32 & 90.35 & 89.91 & -8.89 & -9.41 \\
\bottomrule
\end{tabular}
\endgroup
\end{wraptable}
\textbf{Batch context.}
We examine ASTNN and xASTNN because they pad statement-block sequences to the batch maximum length with zeros, then apply a bidirectional GRU and max pooling without masking padding. Padding can therefore affect program representations according to other programs' lengths in the batch. Table~\ref{tab:rq4_batch_context} compares batch sizes 32 and one with checkpoints, thresholds, pairs, and labels fixed, using the same batch size for original and transformed code. Batch size one evaluates each pair separately, removing other pairs' influence on padding length. On BCB, original F1 falls to 88.83\% and 89.35\%, while transformed F1 remains near 88.6\%. The smaller gaps thus accompany weaker original performance, possibly because batch size changes how zero padding affects representations. On GCJ and OJClone, losses persist and increase at batch size one. Thus, smaller batches do not consistently narrow rewriting losses, and the narrower BCB gap mainly reflects lower original performance rather than improved robustness.
\par\endgroup

\FloatBarrier

\begin{resultbox}
\textbf{Finding 4.}
Behavior-preserving rewriting can weaken discrimination by bringing clone and non-clone scores closer together or changing fixed-threshold decisions while retaining strong ranking ability. The former reflects lost separability, while the latter leaves scope for improvement through threshold adjustment. Input processing also shapes these degradations, linking detector robustness to its inference pipeline.
\end{resultbox}
\FloatBarrier

\section{Analysis and Future Directions}

\subsection{Analysis}

\textbf{Detection effectiveness needs to extend across implementations.}
Clone detection aims to recognize shared functionality across different implementations. RQ1 shows that performance on original code can deteriorate after behavior-preserving rewriting, with detector rankings reversing and previously correct predictions changing even when aggregate scores remain similar. These observations connect robustness to the purpose of clone detection itself. Effective detection requires both correct judgments on the evaluated code and their preservation when implementation details change without altering functionality. Evaluating original and rewritten pairs together therefore tests whether demonstrated effectiveness extends across such changes, while tracking individual predictions reveals which correct judgments are retained.

\textbf{Robustness depends on the interaction between detectors and concrete rewrites.}
RQ2 identifies shared sensitivities across transformation groups, but also shows that detectors within the same representation category respond differently to particular targets and repeated application. RQ3 further shows that alternative constructions at matched locations and edit counts can expose different failures, even when their aggregate losses are similar. Together, these results make concrete construction a substantive part of robustness assessment. Representation categories and transformation families organize the comparison, while variation within them reveals sensitivities that their labels alone do not capture. SCT supports this assessment by organizing transformations at several levels and allowing alternative constructions to be compared under controlled application conditions.

\textbf{Robustness improvements need to address different forms of degradation.}
RQ4 distinguishes reduced score separability from changed decisions that coexist with substantial ranking ability. Threshold adjustment can exploit retained separation, whereas overlapping score distributions call for improvements to the evidence used for discrimination. Input processing adds another dimension. The length comparisons associate estimated limit crossings with more prediction flips, and the batch controls show that processing context changes the measured performance gaps. These findings connect model behavior with the conditions under which code is represented and scored. Improvement strategies should consequently target the observed failure form, assessing whether they preserve discrimination, improve decisions, or stabilize input processing, rather than treating every F1 decline as the same problem.

\subsection{Future Directions}

\textbf{Learning stable discrimination across rewrites.}
The observed increases in both missed clones and accepted non-clones motivate training objectives that preserve clone recognition while maintaining separation from non-clones. SCT variants could support consistency or contrastive objectives that relate a program to its rewrites and retain the distinctions between programs with different functionality \cite{ding2023concord, liu2023contrabert}. Such objectives would target stable discrimination rather than prediction agreement alone. Execution-based approaches offer complementary behavioral signals \cite{li2024prism, wu2020scdetector, mehrotra2021modeling, yuan2022java}. Combining these signals with scalable static representations is a promising direction for extending behavior-aware learning beyond programs for which compilation and execution are readily available.

\textbf{Testing generalization to unseen constructions.}
Sensitivity to alternative constructions raises the question of whether robustness learned from one rewrite transfers to other ways of expressing the same change. Future studies could hold out constructions or operator combinations during training and evaluate them at test time. Comparing these alternatives at common applicable locations and matched edit counts would help characterize transfer across constructions. Reporting coverage, realized edits, and code growth would describe the remaining variation in application conditions, while error overlap would show whether improvements cover the same failures or extend to new ones. This design would distinguish robustness to familiar variants from generalization across implementations.

\textbf{Aligning input processing with robust detection.}
Future detector designs could preserve relevant code as programs grow \cite{zhang2022dietcode} and make program representations independent of other programs in the batch. Candidate approaches include representations that accommodate longer inputs and sequence encoders that explicitly exclude padding from state computation and pooling. Their evaluation should examine original and rewritten performance together, since a smaller gap can result from lower original effectiveness. Comparing predictions and score distributions before and after such changes would establish whether gains come from preserving input information, improving discrimination, or adjusting decision boundaries. This would connect improvements to the complete inference pipeline with the robustness of its clone judgments.

\section{Threats to Validity}
\label{sec:threats}

\textbf{Internal Validity.}
Incorrect rewrites could change program behavior and confound the measured detection losses. SCT checks the contextual requirements of each operator, including the bindings, types, dependencies, and control flow affected by the rewrite, and reparses generated code to reject syntactically invalid candidates. For complete programs, we additionally compile and execute the original and transformed versions under the same conditions, comparing exit status, standard output, and written files as described in Section~\ref{sec:transformation_validation}. No behavioral differences were observed on the tested inputs admitted by these execution checks.

\textbf{External Validity.}
The study covers three benchmarks spanning Java and C, method fragments and complete programs, and code from software projects and programming contests. Evaluating token, tree, graph, and prompt-based detectors broadens the range of approaches examined. Nevertheless, these languages, program sources, and released model configurations represent a subset of practical clone detection settings. We examine sensitivities across the included datasets and methods rather than infer a universal method ranking. Other languages, application domains, model versions, prompts, and reasoning budgets may exhibit different responses and remain targets for broader evaluation.

\textbf{Construct Validity.}
SCT's 51 operators cover selected lexical and structural changes, without exhausting equivalent implementations or representing the distribution of code clones in practice, which is difficult to characterize comprehensively. We organize them by code level, target, and edit category, examine repeated application, and compare alternative constructions at common locations with matched edit counts. These comparisons broaden the assessed variation while retaining explicit application conditions. Clone labels are inherited from the benchmarks, whose operational definitions can differ from pairwise semantic equivalence~\cite{krinke2022harmful}. Preserving pair identities and labels enables comparisons before and after rewriting, but does not independently validate those labels. The study therefore assesses recognition of existing clone relationships under behavior-preserving changes, including robustness relevant to Type-4 detection, rather than accuracy on a separately annotated Type-4 collection.

\textbf{Conclusion Validity.}
Randomness in transformation generation can affect performance losses and the errors uncovered. RQ1 compares results across up to three generation seeds with model checkpoints and evaluation settings fixed to examine whether overall performance declines vary with the generated rewrites. This comparison addresses variation from transformation generation rather than model retraining. To reduce interpretive bias from relying on a single metric, we combine aggregate F1, prediction changes, missed clones and false positives, score separability, and error overlap. These measures distinguish the magnitude of performance losses, changes in failing pairs, and degradation in discrimination, avoiding the interpretation of similar aggregate scores as equivalent detection behavior.

\section{Conclusion}

We characterize the robustness of 14 clone detectors to behavior-preserving rewriting on BCB, GCJ, and OJClone using SCT, a framework of 51 code transformation operators. High benchmark performance does not consistently survive behavior-preserving rewriting. Sensitivity varies with transformation type, application intensity, and concrete construction. Even at matched locations and edit counts, alternative constructions can expose different failures. Degradation takes different forms in score separability and fixed-threshold decisions, leaving different scope for threshold recovery, while input processing also shapes the observed losses. These findings call for extending benchmark effectiveness to stable discrimination across equivalent implementations. Systematic transformation testing provides a concrete way to assess that capability.


\begin{thebibliography}{55}


\ifx \showCODEN    \undefined \def \showCODEN     #1{\unskip}     \fi
\ifx \showISBNx    \undefined \def \showISBNx     #1{\unskip}     \fi
\ifx \showISBNxiii \undefined \def \showISBNxiii  #1{\unskip}     \fi
\ifx \showISSN     \undefined \def \showISSN      #1{\unskip}     \fi
\ifx \showLCCN     \undefined \def \showLCCN      #1{\unskip}     \fi
\ifx \shownote     \undefined \def \shownote      #1{#1}          \fi
\ifx \showarticletitle \undefined \def \showarticletitle #1{#1}   \fi
\ifx \showURL      \undefined \def \showURL       {\relax}        \fi
\providecommand\bibfield[2]{#2}
\providecommand\bibinfo[2]{#2}
\providecommand\natexlab[1]{#1}
\providecommand\showeprint[2][]{arXiv:#2}

\bibitem[{CloneBench}(nd)]%
        {clonebenchrepository}
\bibfield{author}{\bibinfo{person}{{CloneBench}}.}
  \bibinfo{year}{n.d.}\natexlab{}.
\newblock \bibinfo{title}{{BigCloneBench}}.
\newblock \bibinfo{howpublished}{Dataset repository}.
\newblock
\urldef\tempurl%
\url{https://github.com/clonebench/BigCloneBench}
\showURL{%
Retrieved October 23, 2026 from \tempurl}


\bibitem[Dao and Gu(2024)]%
        {dao2024mamba2}
\bibfield{author}{\bibinfo{person}{Tri Dao} {and} \bibinfo{person}{Albert Gu}.}
  \bibinfo{year}{2024}\natexlab{}.
\newblock \showarticletitle{Transformers are {SSM}s: Generalized Models and
  Efficient Algorithms Through Structured State Space Duality}. In
  \bibinfo{booktitle}{\emph{Proceedings of the 41st International Conference on
  Machine Learning}} \emph{(\bibinfo{series}{Proceedings of Machine Learning
  Research}, Vol.~\bibinfo{volume}{235})}. \bibinfo{publisher}{PMLR},
  \bibinfo{pages}{10041--10071}.
\newblock
\urldef\tempurl%
\url{https://proceedings.mlr.press/v235/dao24a.html}
\showURL{%
\tempurl}


\bibitem[{DeepSeek-AI}(2026)]%
        {deepseek2026v41}
\bibfield{author}{\bibinfo{person}{{DeepSeek-AI}}.}
  \bibinfo{year}{2026}\natexlab{}.
\newblock \bibinfo{booktitle}{\emph{{DeepSeek-V4.1-Flash}: Pushing the Limits
  of {KV} Cache Compression}}.
\newblock \bibinfo{type}{Technical Report}. \bibinfo{institution}{DeepSeek-AI}.
\newblock
\urldef\tempurl%
\url{https://huggingface.co/deepseek-ai/DeepSeek-V4.1-Flash/blob/main/DeepSeek_V41_Tech_Report.pdf}
\showURL{%
\tempurl}


\bibitem[Ding et~al\mbox{.}(2023)]%
        {ding2023concord}
\bibfield{author}{\bibinfo{person}{Yangruibo Ding}, \bibinfo{person}{Saikat
  Chakraborty}, \bibinfo{person}{Luca Buratti}, \bibinfo{person}{Saurabh
  Pujar}, \bibinfo{person}{Alessandro Morari}, \bibinfo{person}{Gail Kaiser},
  {and} \bibinfo{person}{Baishakhi Ray}.} \bibinfo{year}{2023}\natexlab{}.
\newblock \showarticletitle{{CONCORD}: Clone-Aware Contrastive Learning for
  Source Code}. In \bibinfo{booktitle}{\emph{Proceedings of the 32nd ACM
  SIGSOFT International Symposium on Software Testing and Analysis}}.
  \bibinfo{publisher}{Association for Computing Machinery},
  \bibinfo{address}{New York, NY, USA}, \bibinfo{pages}{26--38}.
\newblock
\href{https://doi.org/10.1145/3597926.3598035}{doi:\nolinkurl{10.1145/3597926.3598035}}


\bibitem[Dou et~al\mbox{.}(2023)]%
        {dou2023towards}
\bibfield{author}{\bibinfo{person}{Shihan Dou}, \bibinfo{person}{Junjie Shan},
  \bibinfo{person}{Haoxiang Jia}, \bibinfo{person}{Wenhao Deng},
  \bibinfo{person}{Zhiheng Xi}, \bibinfo{person}{Wei He},
  \bibinfo{person}{Yueming Wu}, \bibinfo{person}{Tao Gui},
  \bibinfo{person}{Yang Liu}, {and} \bibinfo{person}{Xuanjing Huang}.}
  \bibinfo{year}{2023}\natexlab{}.
\newblock \bibinfo{title}{Towards Understanding the Capability of Large
  Language Models on Code Clone Detection: A Survey}.
\newblock
\showeprint[arxiv]{2308.01191}~[cs.SE]
\urldef\tempurl%
\url{https://arxiv.org/abs/2308.01191}
\showURL{%
\tempurl}


\bibitem[Dou et~al\mbox{.}(2024)]%
        {dou2024cc2vec}
\bibfield{author}{\bibinfo{person}{Shihan Dou}, \bibinfo{person}{Yueming Wu},
  \bibinfo{person}{Haoxiang Jia}, \bibinfo{person}{Yuhao Zhou},
  \bibinfo{person}{Yan Liu}, {and} \bibinfo{person}{Yang Liu}.}
  \bibinfo{year}{2024}\natexlab{}.
\newblock \showarticletitle{{CC2Vec}: Combining Typed Tokens with Contrastive
  Learning for Effective Code Clone Detection}.
\newblock \bibinfo{journal}{\emph{Proceedings of the ACM on Software
  Engineering}} \bibinfo{volume}{1}, \bibinfo{number}{FSE}
  (\bibinfo{year}{2024}), \bibinfo{pages}{1564--1584}.
\newblock
\href{https://doi.org/10.1145/3660777}{doi:\nolinkurl{10.1145/3660777}}


\bibitem[Dubey et~al\mbox{.}(2024)]%
        {dubey2024llama3}
\bibfield{author}{\bibinfo{person}{Abhimanyu Dubey}, \bibinfo{person}{Abhinav
  Jauhri}, \bibinfo{person}{Abhinav Pandey}, \bibinfo{person}{Abhishek Kadian},
  \bibinfo{person}{Ahmad Al-Dahle}, \bibinfo{person}{Aiesha Letman},
  \bibinfo{person}{Akhil Mathur}, \bibinfo{person}{Alan Schelten},
  \bibinfo{person}{Amy Yang}, \bibinfo{person}{Angela Fan}, {et~al\mbox{.}}}
  \bibinfo{year}{2024}\natexlab{}.
\newblock \bibinfo{title}{The {Llama 3} Herd of Models}.
\newblock
\showeprint[arxiv]{2407.21783v1}~[cs.AI]
\urldef\tempurl%
\url{https://arxiv.org/abs/2407.21783v1}
\showURL{%
\tempurl}


\bibitem[Feng et~al\mbox{.}(2024)]%
        {feng2024machine}
\bibfield{author}{\bibinfo{person}{Siyue Feng}, \bibinfo{person}{Wenqi Suo},
  \bibinfo{person}{Yueming Wu}, \bibinfo{person}{Deqing Zou},
  \bibinfo{person}{Yang Liu}, {and} \bibinfo{person}{Hai Jin}.}
  \bibinfo{year}{2024}\natexlab{}.
\newblock \showarticletitle{Machine Learning is All You Need: A Simple
  Token-Based Approach for Effective Code Clone Detection}. In
  \bibinfo{booktitle}{\emph{2024 IEEE/ACM 46th International Conference on
  Software Engineering (ICSE)}}. \bibinfo{publisher}{IEEE},
  \bibinfo{address}{Piscataway, NJ, USA}, \bibinfo{pages}{2745--2757}.
\newblock
\href{https://doi.org/10.1145/3597503.3639114}{doi:\nolinkurl{10.1145/3597503.3639114}}


\bibitem[Feng et~al\mbox{.}(2020)]%
        {feng2020codebert}
\bibfield{author}{\bibinfo{person}{Zhangyin Feng}, \bibinfo{person}{Daya Guo},
  \bibinfo{person}{Duyu Tang}, \bibinfo{person}{Nan Duan},
  \bibinfo{person}{Xiaocheng Feng}, \bibinfo{person}{Ming Gong},
  \bibinfo{person}{Linjun Shou}, \bibinfo{person}{Bing Qin},
  \bibinfo{person}{Ting Liu}, \bibinfo{person}{Daxin Jiang}, {and}
  \bibinfo{person}{Ming Zhou}.} \bibinfo{year}{2020}\natexlab{}.
\newblock \showarticletitle{{CodeBERT}: A Pre-Trained Model for Programming and
  Natural Languages}. In \bibinfo{booktitle}{\emph{Findings of the Association
  for Computational Linguistics: EMNLP 2020}}. \bibinfo{publisher}{Association
  for Computational Linguistics}, \bibinfo{address}{Online},
  \bibinfo{pages}{1536--1547}.
\newblock
\href{https://doi.org/10.18653/v1/2020.findings-emnlp.139}{doi:\nolinkurl{10.18653/v1/2020.findings-emnlp.139}}


\bibitem[Guo et~al\mbox{.}(2021)]%
        {guo2020graphcodebert}
\bibfield{author}{\bibinfo{person}{Daya Guo}, \bibinfo{person}{Shuo Ren},
  \bibinfo{person}{Shuai Lu}, \bibinfo{person}{Zhangyin Feng},
  \bibinfo{person}{Duyu Tang}, \bibinfo{person}{Shujie Liu},
  \bibinfo{person}{Long Zhou}, \bibinfo{person}{Nan Duan},
  \bibinfo{person}{Alexey Svyatkovskiy}, \bibinfo{person}{Shengyu Fu},
  \bibinfo{person}{Michele Tufano}, \bibinfo{person}{Shao~Kun Deng},
  \bibinfo{person}{Colin Clement}, \bibinfo{person}{Dawn Drain},
  \bibinfo{person}{Neel Sundaresan}, \bibinfo{person}{Jian Yin},
  \bibinfo{person}{Daxin Jiang}, {and} \bibinfo{person}{Ming Zhou}.}
  \bibinfo{year}{2021}\natexlab{}.
\newblock \bibinfo{title}{{GraphCodeBERT}: Pre-Training Code Representations
  with Data Flow}.
\newblock
\showeprint[arxiv]{2009.08366}~[cs.SE]
\urldef\tempurl%
\url{https://arxiv.org/abs/2009.08366}
\showURL{%
\tempurl}


\bibitem[Hou et~al\mbox{.}(2024)]%
        {hou2024llmse}
\bibfield{author}{\bibinfo{person}{Xinyi Hou}, \bibinfo{person}{Yanjie Zhao},
  \bibinfo{person}{Yue Liu}, \bibinfo{person}{Zhou Yang},
  \bibinfo{person}{Kailong Wang}, \bibinfo{person}{Li Li},
  \bibinfo{person}{Xiapu Luo}, \bibinfo{person}{David Lo},
  \bibinfo{person}{John Grundy}, {and} \bibinfo{person}{Haoyu Wang}.}
  \bibinfo{year}{2024}\natexlab{}.
\newblock \showarticletitle{Large Language Models for Software Engineering: A
  Systematic Literature Review}.
\newblock \bibinfo{journal}{\emph{ACM Transactions on Software Engineering and
  Methodology}} \bibinfo{volume}{33}, \bibinfo{number}{8}, Article
  \bibinfo{articleno}{220} (\bibinfo{year}{2024}),
  \bibinfo{numpages}{79}~pages.
\newblock
\href{https://doi.org/10.1145/3695988}{doi:\nolinkurl{10.1145/3695988}}


\bibitem[Jiang et~al\mbox{.}(2007)]%
        {jiang2007deckard}
\bibfield{author}{\bibinfo{person}{Lingxiao Jiang}, \bibinfo{person}{Ghassan
  Misherghi}, \bibinfo{person}{Zhendong Su}, {and} \bibinfo{person}{Stephane
  Glondu}.} \bibinfo{year}{2007}\natexlab{}.
\newblock \showarticletitle{{DECKARD}: Scalable and Accurate Tree-Based
  Detection of Code Clones}. In \bibinfo{booktitle}{\emph{29th International
  Conference on Software Engineering (ICSE'07)}}. \bibinfo{publisher}{IEEE},
  \bibinfo{address}{Piscataway, NJ, USA}, \bibinfo{pages}{96--105}.
\newblock
\href{https://doi.org/10.1109/ICSE.2007.30}{doi:\nolinkurl{10.1109/ICSE.2007.30}}


\bibitem[Juergens et~al\mbox{.}(2009)]%
        {juergens2009code}
\bibfield{author}{\bibinfo{person}{Elmar Juergens}, \bibinfo{person}{Florian
  Deissenboeck}, \bibinfo{person}{Benjamin Hummel}, {and}
  \bibinfo{person}{Stefan Wagner}.} \bibinfo{year}{2009}\natexlab{}.
\newblock \showarticletitle{Do Code Clones Matter?}. In
  \bibinfo{booktitle}{\emph{2009 IEEE 31st International Conference on Software
  Engineering}}. \bibinfo{publisher}{IEEE}, \bibinfo{address}{Piscataway, NJ,
  USA}, \bibinfo{pages}{485--495}.
\newblock
\href{https://doi.org/10.1109/ICSE.2009.5070547}{doi:\nolinkurl{10.1109/ICSE.2009.5070547}}


\bibitem[Kamiya et~al\mbox{.}(2002)]%
        {kamiya2002ccfinder}
\bibfield{author}{\bibinfo{person}{Toshihiro Kamiya}, \bibinfo{person}{Shinji
  Kusumoto}, {and} \bibinfo{person}{Katsuro Inoue}.}
  \bibinfo{year}{2002}\natexlab{}.
\newblock \showarticletitle{{CCFinder}: A Multilinguistic Token-Based Code
  Clone Detection System for Large Scale Source Code}.
\newblock \bibinfo{journal}{\emph{IEEE Transactions on Software Engineering}}
  \bibinfo{volume}{28}, \bibinfo{number}{7} (\bibinfo{year}{2002}),
  \bibinfo{pages}{654--670}.
\newblock
\href{https://doi.org/10.1109/TSE.2002.1019480}{doi:\nolinkurl{10.1109/TSE.2002.1019480}}


\bibitem[Kitsios et~al\mbox{.}(2025)]%
        {kitsios2025detecting}
\bibfield{author}{\bibinfo{person}{Konstantinos Kitsios},
  \bibinfo{person}{Francesco Sovrano}, \bibinfo{person}{Earl~T. Barr}, {and}
  \bibinfo{person}{Alberto Bacchelli}.} \bibinfo{year}{2025}\natexlab{}.
\newblock \showarticletitle{Detecting Semantic Clones of Unseen Functionality}.
  In \bibinfo{booktitle}{\emph{2025 40th IEEE/ACM International Conference on
  Automated Software Engineering (ASE)}}. \bibinfo{publisher}{IEEE},
  \bibinfo{address}{Piscataway, NJ, USA}, \bibinfo{pages}{1312--1324}.
\newblock
\href{https://doi.org/10.1109/ASE63991.2025.00112}{doi:\nolinkurl{10.1109/ASE63991.2025.00112}}


\bibitem[Koschke(2007)]%
        {koschke2007survey}
\bibfield{author}{\bibinfo{person}{Rainer Koschke}.}
  \bibinfo{year}{2007}\natexlab{}.
\newblock \showarticletitle{Survey of Research on Software Clones}. In
  \bibinfo{booktitle}{\emph{Duplication, Redundancy, and Similarity in
  Software}} \emph{(\bibinfo{series}{Dagstuhl Seminar Proceedings},
  Vol.~\bibinfo{volume}{06301})}. \bibinfo{publisher}{Schloss
  Dagstuhl--Leibniz-Zentrum f{\"u}r Informatik}, \bibinfo{address}{Dagstuhl,
  Germany}, \bibinfo{pages}{1--24}.
\newblock
\href{https://doi.org/10.4230/DagSemProc.06301.13}{doi:\nolinkurl{10.4230/DagSemProc.06301.13}}


\bibitem[Krinke and Ragkhitwetsagul(2022)]%
        {krinke2022harmful}
\bibfield{author}{\bibinfo{person}{Jens Krinke} {and} \bibinfo{person}{Chaiyong
  Ragkhitwetsagul}.} \bibinfo{year}{2022}\natexlab{}.
\newblock \showarticletitle{{BigCloneBench} Considered Harmful for Machine
  Learning}. In \bibinfo{booktitle}{\emph{2022 IEEE 16th International Workshop
  on Software Clones (IWSC)}}. \bibinfo{pages}{1--7}.
\newblock
\href{https://doi.org/10.1109/IWSC55060.2022.00008}{doi:\nolinkurl{10.1109/IWSC55060.2022.00008}}


\bibitem[Li et~al\mbox{.}(2024a)]%
        {li2024prism}
\bibfield{author}{\bibinfo{person}{Haoran Li}, \bibinfo{person}{Siqian Wang},
  \bibinfo{person}{Weihong Quan}, \bibinfo{person}{Xiaoli Gong},
  \bibinfo{person}{Huayou Su}, {and} \bibinfo{person}{Jin Zhang}.}
  \bibinfo{year}{2024}\natexlab{a}.
\newblock \showarticletitle{Prism: Decomposing Program Semantics for Code Clone
  Detection through Compilation}. In \bibinfo{booktitle}{\emph{Proceedings of
  the IEEE/ACM 46th International Conference on Software Engineering}}.
  \bibinfo{publisher}{Association for Computing Machinery},
  \bibinfo{address}{New York, NY, USA}, Article \bibinfo{articleno}{217},
  \bibinfo{numpages}{13}~pages.
\newblock
\href{https://doi.org/10.1145/3597503.3639129}{doi:\nolinkurl{10.1145/3597503.3639129}}


\bibitem[Li et~al\mbox{.}(2023)]%
        {li2023zc3}
\bibfield{author}{\bibinfo{person}{Jia Li}, \bibinfo{person}{Chongyang Tao},
  \bibinfo{person}{Zhi Jin}, \bibinfo{person}{Fang Liu}, \bibinfo{person}{Jia
  Li}, {and} \bibinfo{person}{Ge Li}.} \bibinfo{year}{2023}\natexlab{}.
\newblock \showarticletitle{{ZC3}: Zero-Shot Cross-Language Code Clone
  Detection}. In \bibinfo{booktitle}{\emph{2023 38th IEEE/ACM International
  Conference on Automated Software Engineering (ASE)}}.
  \bibinfo{publisher}{IEEE}, \bibinfo{address}{Piscataway, NJ, USA},
  \bibinfo{pages}{875--887}.
\newblock
\href{https://doi.org/10.1109/ASE56229.2023.00210}{doi:\nolinkurl{10.1109/ASE56229.2023.00210}}


\bibitem[Li et~al\mbox{.}(2006)]%
        {li2006cp}
\bibfield{author}{\bibinfo{person}{Zhenmin Li}, \bibinfo{person}{Shan Lu},
  \bibinfo{person}{Suvda Myagmar}, {and} \bibinfo{person}{Yuanyuan Zhou}.}
  \bibinfo{year}{2006}\natexlab{}.
\newblock \showarticletitle{{CP-Miner}: Finding Copy-Paste and Related Bugs in
  Large-Scale Software Code}.
\newblock \bibinfo{journal}{\emph{IEEE Transactions on Software Engineering}}
  \bibinfo{volume}{32}, \bibinfo{number}{3} (\bibinfo{year}{2006}),
  \bibinfo{pages}{176--192}.
\newblock
\href{https://doi.org/10.1109/TSE.2006.28}{doi:\nolinkurl{10.1109/TSE.2006.28}}


\bibitem[Li et~al\mbox{.}(2024b)]%
        {li2024aacegen}
\bibfield{author}{\bibinfo{person}{Zhong Li}, \bibinfo{person}{Chong Zhang},
  \bibinfo{person}{Minxue Pan}, \bibinfo{person}{Tian Zhang}, {and}
  \bibinfo{person}{Xuandong Li}.} \bibinfo{year}{2024}\natexlab{b}.
\newblock \showarticletitle{{AACEGEN}: Attention Guided Adversarial Code
  Example Generation for Deep Code Models}. In
  \bibinfo{booktitle}{\emph{Proceedings of the 39th IEEE/ACM International
  Conference on Automated Software Engineering}}.
  \bibinfo{publisher}{Association for Computing Machinery},
  \bibinfo{address}{New York, NY, USA}, \bibinfo{pages}{1245--1257}.
\newblock
\href{https://doi.org/10.1145/3691620.3695500}{doi:\nolinkurl{10.1145/3691620.3695500}}


\bibitem[Liu et~al\mbox{.}(2025)]%
        {liu2025can}
\bibfield{author}{\bibinfo{person}{Shuo Liu}, \bibinfo{person}{Jacky Keung},
  \bibinfo{person}{Zhen Yang}, \bibinfo{person}{Zhenyu Mao}, {and}
  \bibinfo{person}{Yicheng Sun}.} \bibinfo{year}{2025}\natexlab{}.
\newblock \showarticletitle{Can {Mamba} Be Better? An Experimental Evaluation
  of {Mamba} in Code Intelligence}. In \bibinfo{booktitle}{\emph{2025 40th
  IEEE/ACM International Conference on Automated Software Engineering (ASE)}}.
  \bibinfo{publisher}{IEEE}, \bibinfo{address}{Piscataway, NJ, USA},
  \bibinfo{pages}{1856--1868}.
\newblock
\href{https://doi.org/10.1109/ASE63991.2025.00155}{doi:\nolinkurl{10.1109/ASE63991.2025.00155}}


\bibitem[Liu et~al\mbox{.}(2023)]%
        {liu2023contrabert}
\bibfield{author}{\bibinfo{person}{Shangqing Liu}, \bibinfo{person}{Bozhi Wu},
  \bibinfo{person}{Xiaofei Xie}, \bibinfo{person}{Guozhu Meng}, {and}
  \bibinfo{person}{Yang Liu}.} \bibinfo{year}{2023}\natexlab{}.
\newblock \showarticletitle{{ContraBERT}: Enhancing Code Pre-trained Models via
  Contrastive Learning}. In \bibinfo{booktitle}{\emph{2023 IEEE/ACM 45th
  International Conference on Software Engineering (ICSE)}}.
  \bibinfo{publisher}{IEEE}, \bibinfo{address}{Piscataway, NJ, USA},
  \bibinfo{pages}{2476--2487}.
\newblock
\href{https://doi.org/10.1109/ICSE48619.2023.00207}{doi:\nolinkurl{10.1109/ICSE48619.2023.00207}}


\bibitem[Lu et~al\mbox{.}(2021)]%
        {lu2021codexglue}
\bibfield{author}{\bibinfo{person}{Shuai Lu}, \bibinfo{person}{Daya Guo},
  \bibinfo{person}{Shuo Ren}, \bibinfo{person}{Junjie Huang},
  \bibinfo{person}{Alexey Svyatkovskiy}, \bibinfo{person}{Ambrosio Blanco},
  \bibinfo{person}{Colin Clement}, \bibinfo{person}{Dawn Drain},
  \bibinfo{person}{Daxin Jiang}, \bibinfo{person}{Duyu Tang},
  \bibinfo{person}{Ge Li}, \bibinfo{person}{Lidong Zhou},
  \bibinfo{person}{Linjun Shou}, \bibinfo{person}{Long Zhou},
  \bibinfo{person}{Michele Tufano}, \bibinfo{person}{Ming Gong},
  \bibinfo{person}{Ming Zhou}, \bibinfo{person}{Nan Duan},
  \bibinfo{person}{Neel Sundaresan}, \bibinfo{person}{Shao~Kun Deng},
  \bibinfo{person}{Shengyu Fu}, {and} \bibinfo{person}{Shujie Liu}.}
  \bibinfo{year}{2021}\natexlab{}.
\newblock \showarticletitle{{CodeXGLUE}: A Machine Learning Benchmark Dataset
  for Code Understanding and Generation}. In
  \bibinfo{booktitle}{\emph{Proceedings of the Neural Information Processing
  Systems Track on Datasets and Benchmarks}}, Vol.~\bibinfo{volume}{1}.
\newblock
\urldef\tempurl%
\url{https://datasets-benchmarks-proceedings.neurips.cc/paper_files/paper/2021/hash/c16a5320fa475530d9583c34fd356ef5-Abstract-round1.html}
\showURL{%
\tempurl}


\bibitem[Mehrotra et~al\mbox{.}(2022)]%
        {mehrotra2021modeling}
\bibfield{author}{\bibinfo{person}{Nikita Mehrotra}, \bibinfo{person}{Navdha
  Agarwal}, \bibinfo{person}{Piyush Gupta}, \bibinfo{person}{Saket Anand},
  \bibinfo{person}{David Lo}, {and} \bibinfo{person}{Rahul Purandare}.}
  \bibinfo{year}{2022}\natexlab{}.
\newblock \showarticletitle{Modeling Functional Similarity in Source Code with
  Graph-Based Siamese Networks}.
\newblock \bibinfo{journal}{\emph{IEEE Transactions on Software Engineering}}
  \bibinfo{volume}{48}, \bibinfo{number}{10} (\bibinfo{year}{2022}),
  \bibinfo{pages}{3771--3789}.
\newblock
\href{https://doi.org/10.1109/TSE.2021.3105556}{doi:\nolinkurl{10.1109/TSE.2021.3105556}}


\bibitem[Mehrotra et~al\mbox{.}(2023)]%
        {mehrotra2023rubhus}
\bibfield{author}{\bibinfo{person}{Nikita Mehrotra}, \bibinfo{person}{Akash
  Sharma}, \bibinfo{person}{Anmol Jindal}, {and} \bibinfo{person}{Rahul
  Purandare}.} \bibinfo{year}{2023}\natexlab{}.
\newblock \showarticletitle{Improving Cross-Language Code Clone Detection via
  Code Representation Learning and Graph Neural Networks}.
\newblock \bibinfo{journal}{\emph{IEEE Transactions on Software Engineering}}
  \bibinfo{volume}{49}, \bibinfo{number}{11} (\bibinfo{year}{2023}),
  \bibinfo{pages}{4846--4868}.
\newblock
\href{https://doi.org/10.1109/TSE.2023.3311796}{doi:\nolinkurl{10.1109/TSE.2023.3311796}}


\bibitem[{Microsoft}(nd)]%
        {microsoftcodexgluebcb}
\bibfield{author}{\bibinfo{person}{{Microsoft}}.}
  \bibinfo{year}{n.d.}\natexlab{}.
\newblock \bibinfo{title}{{CodeXGLUE}: Clone Detection on {BigCloneBench}}.
\newblock \bibinfo{howpublished}{Dataset and evaluation files}.
\newblock
\urldef\tempurl%
\url{https://github.com/microsoft/CodeXGLUE/tree/main/Code-Code/Clone-detection-BigCloneBench}
\showURL{%
Retrieved October 23, 2026 from \tempurl}


\bibitem[Mou et~al\mbox{.}(2016)]%
        {mou2016convolutional}
\bibfield{author}{\bibinfo{person}{Lili Mou}, \bibinfo{person}{Ge Li},
  \bibinfo{person}{Lu Zhang}, \bibinfo{person}{Tao Wang}, {and}
  \bibinfo{person}{Zhi Jin}.} \bibinfo{year}{2016}\natexlab{}.
\newblock \showarticletitle{Convolutional Neural Networks over Tree Structures
  for Programming Language Processing}. In
  \bibinfo{booktitle}{\emph{Proceedings of the AAAI Conference on Artificial
  Intelligence}}, Vol.~\bibinfo{volume}{30}. \bibinfo{publisher}{AAAI Press},
  \bibinfo{address}{Palo Alto, CA, USA}, \bibinfo{pages}{1287--1293}.
\newblock
\href{https://doi.org/10.1609/aaai.v30i1.10139}{doi:\nolinkurl{10.1609/aaai.v30i1.10139}}


\bibitem[Moumoula et~al\mbox{.}(2025)]%
        {moumoula2025struggles}
\bibfield{author}{\bibinfo{person}{Micheline~B{\'e}n{\'e}dicte Moumoula},
  \bibinfo{person}{Abdoul~Kader Kabor{\'e}}, \bibinfo{person}{Jacques Klein},
  {and} \bibinfo{person}{Tegawend{\'e}~F. Bissyand{\'e}}.}
  \bibinfo{year}{2025}\natexlab{}.
\newblock \showarticletitle{The Struggles of {LLMs} in Cross-Lingual Code Clone
  Detection}.
\newblock \bibinfo{journal}{\emph{Proceedings of the ACM on Software
  Engineering}} \bibinfo{volume}{2}, \bibinfo{number}{FSE}, Article
  \bibinfo{articleno}{FSE047} (\bibinfo{year}{2025}),
  \bibinfo{numpages}{23}~pages.
\newblock
\href{https://doi.org/10.1145/3715764}{doi:\nolinkurl{10.1145/3715764}}


\bibitem[{OpenAI}(2025)]%
        {openai2025gptoss}
\bibfield{author}{\bibinfo{person}{{OpenAI}}.} \bibinfo{year}{2025}\natexlab{}.
\newblock \bibinfo{title}{{GPT-OSS-20B}}.
\newblock \bibinfo{howpublished}{Official model card}.
\newblock
\urldef\tempurl%
\url{https://huggingface.co/openai/gpt-oss-20b}
\showURL{%
Retrieved October 30, 2026 from \tempurl}


\bibitem[Panichella(2025)]%
        {panichella2025morph}
\bibfield{author}{\bibinfo{person}{Annibale Panichella}.}
  \bibinfo{year}{2025}\natexlab{}.
\newblock \showarticletitle{Metamorphic-Based Many-Objective Distillation of
  {LLMs} for Code-Related Tasks}. In \bibinfo{booktitle}{\emph{2025 IEEE/ACM
  47th International Conference on Software Engineering (ICSE)}}.
  \bibinfo{publisher}{IEEE}, \bibinfo{address}{Piscataway, NJ, USA},
  \bibinfo{pages}{1001--1013}.
\newblock
\href{https://doi.org/10.1109/ICSE55347.2025.00230}{doi:\nolinkurl{10.1109/ICSE55347.2025.00230}}


\bibitem[Roy and Cordy(2009)]%
        {roy2009mutation}
\bibfield{author}{\bibinfo{person}{Chanchal~K. Roy} {and}
  \bibinfo{person}{James~R. Cordy}.} \bibinfo{year}{2009}\natexlab{}.
\newblock \showarticletitle{A Mutation/Injection-Based Automatic Framework for
  Evaluating Code Clone Detection Tools}. In \bibinfo{booktitle}{\emph{2009
  International Conference on Software Testing Verification and Validation
  Workshops}}. \bibinfo{publisher}{IEEE}, \bibinfo{address}{Piscataway, NJ,
  USA}, \bibinfo{pages}{157--166}.
\newblock
\href{https://doi.org/10.1109/ICSTW.2009.18}{doi:\nolinkurl{10.1109/ICSTW.2009.18}}


\bibitem[Saini et~al\mbox{.}(2018)]%
        {saini2018oreo}
\bibfield{author}{\bibinfo{person}{Vaibhav Saini}, \bibinfo{person}{Farima
  Farmahinifarahani}, \bibinfo{person}{Yadong Lu}, \bibinfo{person}{Pierre
  Baldi}, {and} \bibinfo{person}{Cristina~V. Lopes}.}
  \bibinfo{year}{2018}\natexlab{}.
\newblock \showarticletitle{Oreo: Detection of Clones in the Twilight Zone}. In
  \bibinfo{booktitle}{\emph{Proceedings of the 2018 26th ACM Joint Meeting on
  European Software Engineering Conference and Symposium on the Foundations of
  Software Engineering}}. \bibinfo{publisher}{Association for Computing
  Machinery}, \bibinfo{address}{New York, NY, USA}, \bibinfo{pages}{354--365}.
\newblock
\href{https://doi.org/10.1145/3236024.3236026}{doi:\nolinkurl{10.1145/3236024.3236026}}


\bibitem[Sajnani et~al\mbox{.}(2016)]%
        {sajnani2016sourcerercc}
\bibfield{author}{\bibinfo{person}{Hitesh Sajnani}, \bibinfo{person}{Vaibhav
  Saini}, \bibinfo{person}{Jeffrey Svajlenko}, \bibinfo{person}{Chanchal~K.
  Roy}, {and} \bibinfo{person}{Cristina~V. Lopes}.}
  \bibinfo{year}{2016}\natexlab{}.
\newblock \showarticletitle{{SourcererCC}: Scaling Code Clone Detection to
  Big-Code}. In \bibinfo{booktitle}{\emph{Proceedings of the 38th International
  Conference on Software Engineering}}. \bibinfo{publisher}{Association for
  Computing Machinery}, \bibinfo{address}{New York, NY, USA},
  \bibinfo{pages}{1157--1168}.
\newblock
\href{https://doi.org/10.1145/2884781.2884877}{doi:\nolinkurl{10.1145/2884781.2884877}}


\bibitem[Svajlenko et~al\mbox{.}(2014)]%
        {svajlenko2014towards}
\bibfield{author}{\bibinfo{person}{Jeffrey Svajlenko},
  \bibinfo{person}{Judith~F. Islam}, \bibinfo{person}{Iman Keivanloo},
  \bibinfo{person}{Chanchal~K. Roy}, {and} \bibinfo{person}{Mohammad~Mamun
  Mia}.} \bibinfo{year}{2014}\natexlab{}.
\newblock \showarticletitle{Towards a Big Data Curated Benchmark of
  Inter-Project Code Clones}. In \bibinfo{booktitle}{\emph{2014 IEEE
  International Conference on Software Maintenance and Evolution}}.
  \bibinfo{publisher}{IEEE}, \bibinfo{address}{Piscataway, NJ, USA},
  \bibinfo{pages}{476--480}.
\newblock
\href{https://doi.org/10.1109/ICSME.2014.77}{doi:\nolinkurl{10.1109/ICSME.2014.77}}


\bibitem[Tian et~al\mbox{.}(2023)]%
        {tian2023coda}
\bibfield{author}{\bibinfo{person}{Zhao Tian}, \bibinfo{person}{Junjie Chen},
  {and} \bibinfo{person}{Zhi Jin}.} \bibinfo{year}{2023}\natexlab{}.
\newblock \showarticletitle{Code Difference Guided Adversarial Example
  Generation for Deep Code Models}. In \bibinfo{booktitle}{\emph{2023 38th
  IEEE/ACM International Conference on Automated Software Engineering (ASE)}}.
  \bibinfo{publisher}{IEEE}, \bibinfo{address}{Piscataway, NJ, USA},
  \bibinfo{pages}{850--862}.
\newblock
\href{https://doi.org/10.1109/ASE56229.2023.00149}{doi:\nolinkurl{10.1109/ASE56229.2023.00149}}


\bibitem[{TypeSafe AI}(2026)]%
        {typesafe2026jev}
\bibfield{author}{\bibinfo{person}{{TypeSafe AI}}.}
  \bibinfo{year}{2026}\natexlab{}.
\newblock \bibinfo{title}{Models}.
\newblock \bibinfo{howpublished}{Official Jev model documentation}.
\newblock
\urldef\tempurl%
\url{https://docs.typesafe.ai/models}
\showURL{%
Retrieved October 23, 2026 from \tempurl}


\bibitem[Wang et~al\mbox{.}(2020)]%
        {wang2020detecting}
\bibfield{author}{\bibinfo{person}{Wenhan Wang}, \bibinfo{person}{Ge Li},
  \bibinfo{person}{Bo Ma}, \bibinfo{person}{Xin Xia}, {and}
  \bibinfo{person}{Zhi Jin}.} \bibinfo{year}{2020}\natexlab{}.
\newblock \showarticletitle{Detecting Code Clones with Graph Neural Network and
  Flow-Augmented Abstract Syntax Tree}. In \bibinfo{booktitle}{\emph{2020 IEEE
  27th International Conference on Software Analysis, Evolution and
  Reengineering (SANER)}}. \bibinfo{publisher}{IEEE},
  \bibinfo{address}{Piscataway, NJ, USA}, \bibinfo{pages}{261--271}.
\newblock
\href{https://doi.org/10.1109/SANER48275.2020.9054857}{doi:\nolinkurl{10.1109/SANER48275.2020.9054857}}


\bibitem[Wang et~al\mbox{.}(2023)]%
        {wang2023comparison}
\bibfield{author}{\bibinfo{person}{Yuekun Wang}, \bibinfo{person}{Yuhang Ye},
  \bibinfo{person}{Yueming Wu}, \bibinfo{person}{Weiwei Zhang},
  \bibinfo{person}{Yinxing Xue}, {and} \bibinfo{person}{Yang Liu}.}
  \bibinfo{year}{2023}\natexlab{}.
\newblock \showarticletitle{Comparison and Evaluation of Clone Detection
  Techniques with Different Code Representations}. In
  \bibinfo{booktitle}{\emph{2023 IEEE/ACM 45th International Conference on
  Software Engineering (ICSE)}}. \bibinfo{publisher}{IEEE},
  \bibinfo{address}{Piscataway, NJ, USA}, \bibinfo{pages}{332--344}.
\newblock
\href{https://doi.org/10.1109/ICSE48619.2023.00039}{doi:\nolinkurl{10.1109/ICSE48619.2023.00039}}


\bibitem[Wei and Li(2017)]%
        {wei2017supervised}
\bibfield{author}{\bibinfo{person}{Hui-Hui Wei} {and} \bibinfo{person}{Ming
  Li}.} \bibinfo{year}{2017}\natexlab{}.
\newblock \showarticletitle{Supervised Deep Features for Software Functional
  Clone Detection by Exploiting Lexical and Syntactical Information in Source
  Code}. In \bibinfo{booktitle}{\emph{Proceedings of the 26th International
  Joint Conference on Artificial Intelligence}}.
  \bibinfo{publisher}{International Joint Conferences on Artificial
  Intelligence Organization}, \bibinfo{address}{California, USA},
  \bibinfo{pages}{3034--3040}.
\newblock
\href{https://doi.org/10.24963/ijcai.2017/423}{doi:\nolinkurl{10.24963/ijcai.2017/423}}


\bibitem[Wu et~al\mbox{.}(2021)]%
        {wu2020scdetector}
\bibfield{author}{\bibinfo{person}{Yueming Wu}, \bibinfo{person}{Deqing Zou},
  \bibinfo{person}{Shihan Dou}, \bibinfo{person}{Siru Yang},
  \bibinfo{person}{Wei Yang}, \bibinfo{person}{Feng Cheng},
  \bibinfo{person}{Hong Liang}, {and} \bibinfo{person}{Hai Jin}.}
  \bibinfo{year}{2021}\natexlab{}.
\newblock \showarticletitle{{SCDetector}: Software Functional Clone Detection
  Based on Semantic Tokens Analysis}. In \bibinfo{booktitle}{\emph{Proceedings
  of the 35th IEEE/ACM International Conference on Automated Software
  Engineering}}. \bibinfo{publisher}{Association for Computing Machinery},
  \bibinfo{address}{New York, NY, USA}, \bibinfo{pages}{821--833}.
\newblock
\href{https://doi.org/10.1145/3324884.3416562}{doi:\nolinkurl{10.1145/3324884.3416562}}


\bibitem[Xu et~al\mbox{.}(2024)]%
        {xu2024dsfm}
\bibfield{author}{\bibinfo{person}{Zhiwei Xu}, \bibinfo{person}{Shaohua Qiang},
  \bibinfo{person}{Dinghong Song}, \bibinfo{person}{Min Zhou},
  \bibinfo{person}{Hai Wan}, \bibinfo{person}{Xibin Zhao},
  \bibinfo{person}{Ping Luo}, {and} \bibinfo{person}{Hongyu Zhang}.}
  \bibinfo{year}{2024}\natexlab{}.
\newblock \showarticletitle{{DSFM}: Enhancing Functional Code Clone Detection
  with Deep Subtree Interactions}. In \bibinfo{booktitle}{\emph{2024 IEEE/ACM
  46th International Conference on Software Engineering (ICSE)}}.
  \bibinfo{publisher}{IEEE}, \bibinfo{address}{Piscataway, NJ, USA},
  \bibinfo{pages}{2733--2744}.
\newblock
\href{https://doi.org/10.1145/3597503.3639215}{doi:\nolinkurl{10.1145/3597503.3639215}}


\bibitem[Xu et~al\mbox{.}(2023)]%
        {xu2023xastnn}
\bibfield{author}{\bibinfo{person}{Zhiwei Xu}, \bibinfo{person}{Min Zhou},
  \bibinfo{person}{Xibin Zhao}, \bibinfo{person}{Yang Chen},
  \bibinfo{person}{Xi Cheng}, {and} \bibinfo{person}{Hongyu Zhang}.}
  \bibinfo{year}{2023}\natexlab{}.
\newblock \showarticletitle{{xASTNN}: Improved Code Representations for
  Industrial Practice}. In \bibinfo{booktitle}{\emph{Proceedings of the 31st
  ACM Joint European Software Engineering Conference and Symposium on the
  Foundations of Software Engineering}}. \bibinfo{publisher}{Association for
  Computing Machinery}, \bibinfo{address}{New York, NY, USA},
  \bibinfo{pages}{1727--1738}.
\newblock
\href{https://doi.org/10.1145/3611643.3613869}{doi:\nolinkurl{10.1145/3611643.3613869}}


\bibitem[Yang et~al\mbox{.}(2022)]%
        {yang2022natural}
\bibfield{author}{\bibinfo{person}{Zhou Yang}, \bibinfo{person}{Jieke Shi},
  \bibinfo{person}{Junda He}, {and} \bibinfo{person}{David Lo}.}
  \bibinfo{year}{2022}\natexlab{}.
\newblock \showarticletitle{Natural Attack for Pre-trained Models of Code}. In
  \bibinfo{booktitle}{\emph{Proceedings of the 44th International Conference on
  Software Engineering}}. \bibinfo{publisher}{Association for Computing
  Machinery}, \bibinfo{address}{New York, NY, USA},
  \bibinfo{pages}{1482--1493}.
\newblock
\href{https://doi.org/10.1145/3510003.3510146}{doi:\nolinkurl{10.1145/3510003.3510146}}


\bibitem[Yu et~al\mbox{.}(2025)]%
        {yu2025multiple}
\bibfield{author}{\bibinfo{person}{Tianchen Yu}, \bibinfo{person}{Li Yuan},
  \bibinfo{person}{Liannan Lin}, {and} \bibinfo{person}{Hongkui He}.}
  \bibinfo{year}{2025}\natexlab{}.
\newblock \showarticletitle{A Multiple Representation Transformer with
  Optimized Abstract Syntax Tree for Efficient Code Clone Detection}. In
  \bibinfo{booktitle}{\emph{2025 IEEE/ACM 47th International Conference on
  Software Engineering (ICSE)}}. \bibinfo{publisher}{IEEE},
  \bibinfo{address}{Piscataway, NJ, USA}, \bibinfo{pages}{281--293}.
\newblock
\href{https://doi.org/10.1109/ICSE55347.2025.00050}{doi:\nolinkurl{10.1109/ICSE55347.2025.00050}}


\bibitem[Yuan et~al\mbox{.}(2023)]%
        {yuan2022java}
\bibfield{author}{\bibinfo{person}{Dawei Yuan}, \bibinfo{person}{Sen Fang},
  \bibinfo{person}{Tao Zhang}, \bibinfo{person}{Zhou Xu}, {and}
  \bibinfo{person}{Xiapu Luo}.} \bibinfo{year}{2023}\natexlab{}.
\newblock \showarticletitle{Java Code Clone Detection by Exploiting Semantic
  and Syntax Information from Intermediate Code-Based Graph}.
\newblock \bibinfo{journal}{\emph{IEEE Transactions on Reliability}}
  \bibinfo{volume}{72}, \bibinfo{number}{2} (\bibinfo{year}{2023}),
  \bibinfo{pages}{511--526}.
\newblock
\href{https://doi.org/10.1109/TR.2022.3176922}{doi:\nolinkurl{10.1109/TR.2022.3176922}}


\bibitem[Zhang et~al\mbox{.}(2022a)]%
        {zhang2022carrot}
\bibfield{author}{\bibinfo{person}{Huangzhao Zhang}, \bibinfo{person}{Zhiyi
  Fu}, \bibinfo{person}{Ge Li}, \bibinfo{person}{Lei Ma},
  \bibinfo{person}{Zhehao Zhao}, \bibinfo{person}{Hua'an Yang},
  \bibinfo{person}{Yizhe Sun}, \bibinfo{person}{Yang Liu}, {and}
  \bibinfo{person}{Zhi Jin}.} \bibinfo{year}{2022}\natexlab{a}.
\newblock \showarticletitle{Towards Robustness of Deep Program Processing
  Models---Detection, Estimation, and Enhancement}.
\newblock \bibinfo{journal}{\emph{ACM Transactions on Software Engineering and
  Methodology}} \bibinfo{volume}{31}, \bibinfo{number}{3}, Article
  \bibinfo{articleno}{50} (\bibinfo{year}{2022}), \bibinfo{numpages}{40}~pages.
\newblock
\href{https://doi.org/10.1145/3511887}{doi:\nolinkurl{10.1145/3511887}}


\bibitem[Zhang et~al\mbox{.}(2019)]%
        {zhang2019novel}
\bibfield{author}{\bibinfo{person}{Jian Zhang}, \bibinfo{person}{Xu Wang},
  \bibinfo{person}{Hongyu Zhang}, \bibinfo{person}{Hailong Sun},
  \bibinfo{person}{Kaixuan Wang}, {and} \bibinfo{person}{Xudong Liu}.}
  \bibinfo{year}{2019}\natexlab{}.
\newblock \showarticletitle{A Novel Neural Source Code Representation Based on
  Abstract Syntax Tree}. In \bibinfo{booktitle}{\emph{2019 IEEE/ACM 41st
  International Conference on Software Engineering (ICSE)}}.
  \bibinfo{publisher}{IEEE}, \bibinfo{address}{Piscataway, NJ, USA},
  \bibinfo{pages}{783--794}.
\newblock
\href{https://doi.org/10.1109/ICSE.2019.00086}{doi:\nolinkurl{10.1109/ICSE.2019.00086}}


\bibitem[Zhang et~al\mbox{.}(nd)]%
        {astnnartifact}
\bibfield{author}{\bibinfo{person}{Jian Zhang}, \bibinfo{person}{Xu Wang},
  \bibinfo{person}{Hongyu Zhang}, \bibinfo{person}{Hailong Sun},
  \bibinfo{person}{Kaixuan Wang}, {and} \bibinfo{person}{Xudong Liu}.}
  \bibinfo{year}{n.d.}\natexlab{}.
\newblock \bibinfo{title}{{ASTNN}: Source Code and Experimental Data}.
\newblock \bibinfo{howpublished}{Author implementation and dataset release}.
\newblock
\urldef\tempurl%
\url{https://github.com/zhangj111/astnn}
\showURL{%
Retrieved October 23, 2026 from \tempurl}


\bibitem[Zhang et~al\mbox{.}(2023)]%
        {zhang2023clonegen}
\bibfield{author}{\bibinfo{person}{Weiwei Zhang}, \bibinfo{person}{Shengjian
  Guo}, \bibinfo{person}{Hongyu Zhang}, \bibinfo{person}{Yulei Sui},
  \bibinfo{person}{Yinxing Xue}, {and} \bibinfo{person}{Yun Xu}.}
  \bibinfo{year}{2023}\natexlab{}.
\newblock \showarticletitle{Challenging Machine Learning-Based Clone Detectors
  via Semantic-Preserving Code Transformations}.
\newblock \bibinfo{journal}{\emph{IEEE Transactions on Software Engineering}}
  \bibinfo{volume}{49}, \bibinfo{number}{5} (\bibinfo{year}{2023}),
  \bibinfo{pages}{3052--3070}.
\newblock
\href{https://doi.org/10.1109/TSE.2023.3240118}{doi:\nolinkurl{10.1109/TSE.2023.3240118}}


\bibitem[Zhang et~al\mbox{.}(2022b)]%
        {zhang2022dietcode}
\bibfield{author}{\bibinfo{person}{Zhaowei Zhang}, \bibinfo{person}{Hongyu
  Zhang}, \bibinfo{person}{Beijun Shen}, {and} \bibinfo{person}{Xiaodong Gu}.}
  \bibinfo{year}{2022}\natexlab{b}.
\newblock \showarticletitle{Diet Code Is Healthy: Simplifying Programs for
  Pre-trained Models of Code}. In \bibinfo{booktitle}{\emph{Proceedings of the
  30th ACM Joint European Software Engineering Conference and Symposium on the
  Foundations of Software Engineering}}. \bibinfo{publisher}{Association for
  Computing Machinery}, \bibinfo{address}{New York, NY, USA},
  \bibinfo{pages}{1073--1084}.
\newblock
\href{https://doi.org/10.1145/3540250.3549094}{doi:\nolinkurl{10.1145/3540250.3549094}}


\bibitem[Zhao and Huang(2018)]%
        {zhao2018deepsim}
\bibfield{author}{\bibinfo{person}{Gang Zhao} {and} \bibinfo{person}{Jeff
  Huang}.} \bibinfo{year}{2018}\natexlab{}.
\newblock \showarticletitle{{DeepSim}: Deep Learning Code Functional
  Similarity}. In \bibinfo{booktitle}{\emph{Proceedings of the 2018 26th ACM
  Joint Meeting on European Software Engineering Conference and Symposium on
  the Foundations of Software Engineering}}. \bibinfo{publisher}{Association
  for Computing Machinery}, \bibinfo{address}{New York, NY, USA},
  \bibinfo{pages}{141--151}.
\newblock
\href{https://doi.org/10.1145/3236024.3236068}{doi:\nolinkurl{10.1145/3236024.3236068}}


\bibitem[Zhao and Huang(nd)]%
        {deepsimartifact}
\bibfield{author}{\bibinfo{person}{Gang Zhao} {and} \bibinfo{person}{Jeff
  Huang}.} \bibinfo{year}{n.d.}\natexlab{}.
\newblock \bibinfo{title}{{DeepSim}}.
\newblock \bibinfo{howpublished}{Source code and GCJ experimental data}.
\newblock
\urldef\tempurl%
\url{https://github.com/parasol-aser/deepsim}
\showURL{%
Retrieved October 23, 2026 from \tempurl}


\bibitem[Zhu et~al\mbox{.}(2025)]%
        {zhu2025empirical}
\bibfield{author}{\bibinfo{person}{Wenqing Zhu}, \bibinfo{person}{Norihiro
  Yoshida}, \bibinfo{person}{Eunjong Choi}, \bibinfo{person}{Yutaka Matsubara},
  {and} \bibinfo{person}{Hiroaki Takada}.} \bibinfo{year}{2025}\natexlab{}.
\newblock \bibinfo{title}{An Empirical Study of {LLM}-Based Code Clone
  Detection}.
\newblock
\showeprint[arxiv]{2511.01176}~[cs.SE]
\urldef\tempurl%
\url{https://arxiv.org/abs/2511.01176}
\showURL{%
\tempurl}


\bibitem[Zou et~al\mbox{.}(2020)]%
        {zou2020ccgraph}
\bibfield{author}{\bibinfo{person}{Yue Zou}, \bibinfo{person}{Bihuan Ban},
  \bibinfo{person}{Yinxing Xue}, {and} \bibinfo{person}{Yun Xu}.}
  \bibinfo{year}{2020}\natexlab{}.
\newblock \showarticletitle{{CCGraph}: A {PDG}-Based Code Clone Detector with
  Approximate Graph Matching}. In \bibinfo{booktitle}{\emph{Proceedings of the
  35th IEEE/ACM International Conference on Automated Software Engineering}}.
  \bibinfo{publisher}{Association for Computing Machinery},
  \bibinfo{address}{New York, NY, USA}, \bibinfo{pages}{931--942}.
\newblock
\href{https://doi.org/10.1145/3324884.3416541}{doi:\nolinkurl{10.1145/3324884.3416541}}


\end{thebibliography}
\end{document}